\documentclass[12pt]{article}

\usepackage{amssymb}

\makeatletter
 \@addtoreset{equation}{section}
 
\makeatother

\newcommand{\maru}{\mbox{\tiny$\stackrel{\circ}{\scriptstyle\circ}$}}
\newcommand{\mm}{\mathrm{m}}
\newcommand{\mn}{\mathrm{n}}

\newcommand{\pdot}{{\displaystyle{\raisebox{-1.5ex}[0.25ex]{$\cdot$}
     \atop\raisebox{0.6ex}[0.25ex]{$\scriptstyle (p)$}}}}

\begin{document}

%
%
\begin{titlepage}

\renewcommand{\thefootnote}{\fnsymbol{footnote}}

\begin{flushright}
      \normalsize    April 2001\\
                     OU-HET 383\\ 
                     hep-th/0104243
\end{flushright}

\vspace{2ex}

\begin{center}
  {\Large \bf $p-p^{\prime}$ System with $B$ Field and\\[1ex]
  Projection Operator Noncommutative Solitons}
\end{center}

\vspace{2ex}

\begin{center}
    Koichi {\sc Murakami}\footnote{
      e-mail: murakami@het.phys.sci.osaka-u.ac.jp}
\end{center}

\vspace{2ex}

\begin{center}
      \it  Department of Physics,
        Graduate School of Science, Osaka University,\\
        Toyonaka, Osaka 560-0043, Japan
\end{center}

\vspace{2ex}


\begin{abstract}
We study the system of the D$p^{\prime}$-brane
with D$p$-brane ($p<p^{\prime}$) inside,
in the case where $B_{ij}$ field is a nonvanishing constant.
In order to understand how the D$p$-brane is
viewed from the D$p^{\prime}$-brane worldvolume theory,
we investigate the process in which the D$p$-brane is probed
with $p^{\prime}$-$p^{\prime}$ open string.
We calculate the scattering amplitudes among $p$-$p^{\prime}$
open strings and $p^{\prime}$-$p^{\prime}$ open strings
and show that not only the Weyl transform of the projection
operator onto the ground state
but also those onto higher excited states
emerge as multiplicative factors of the amplitudes.

\end{abstract}

\vfill

\setcounter{footnote}{0}
\renewcommand{\thefootnote}{\arabic{footnote}}

\end{titlepage}

\section{Introduction}                                       

String theory with a constant NS-NS two-form $B$ field background
has several interesting features \cite{FT}\cite{CLNY}\cite{ACNY}.
Among others, it has been found out that
when D-branes accompany this system
the worldvolume of the D-branes
becomes noncommutative \cite{CDS}\cite{DH}\cite{CH}\cite{AJ}\cite{SW}.

In this paper we study the $p$-$p^{\prime}$ system:
the system of a D$p^{\prime}$-brane into which a D$p$-brane $(p<p^{\prime})$
is embedded, with a constant $B$ field background.
We would like to approach this system
in terms of  perturbative string theory.
Several aspects of the $p$-$p^{\prime}$ system have been analyzed
in the framework of  string theory
\cite{AH}\cite{GNS}\cite{Garousi}\cite{FGRS}.
In the present paper, along the line of the analysis of \cite{CIMM2},
we investigate how the lower dimensional D-brane (i.e.\ D$p$-brane)
is viewed from the higher dimensional D-brane
(i.e.\ D$p^{\prime}$-brane). For this purpose,
we examine the process in which the D$p$-brane
is probed with $p^{\prime}$-$p^{\prime}$ open strings.
In the perturbative string theory, D-branes couple to the strings
through the open strings attaching on them.
Among the open strings ending on the D$p$-brane,
the $p$-$p^{\prime}$ open string
directly couples to the $p^{\prime}$-$p^{\prime}$ open string.
Therefore the consideration of the process probing the D$p$-brane
with the $p^{\prime}$-$p^{\prime}$ string
leads us to evaluate
scattering amplitudes among $p$-$p^{\prime}$ open strings and
$p^{\prime}$-$p^{\prime}$ open strings.

In refs.~\cite{CIMM2} and \cite{CIMM3}, from the evaluation of
the $N$-point scattering amplitudes consisting of two
ground state vertex operators of the $p$-$p^{\prime}$ open string
and $(N-2)$ gauge field vertex operators of
the $p^{\prime}$-$p^{\prime}$ open string,
it has been concluded that in the zero slope limit
the form factor of the D$p$-brane
becomes precisely the Weyl transform of the projection operator
$|0\rangle \langle 0|$, which is a classical solution
obtained by \cite{GMS} in the noncommutative field theory.
This fact suggests that the D$p$-brane within the noncommutative
D$p^{\prime}$-brane
should be described as the noncommutative soliton.
Consistent results  with this observation
have been obtained from the analyses in terms of field theory
\cite{DMR}\cite{HKLM}\cite{EW}\cite{Pol}\cite{Bak}\cite{AGMS}.
Such coincidence leads us to the question whether
it is possible to read from the string scattering amplitudes
the projection operators $|\mm\rangle\langle \mm|$ ($\mm=1,2,\ldots$)
besides that onto
the harmonic oscillator ground state $|0\rangle\langle 0|$.

One of the remarkable properties of the $p$-$p^{\prime}$ system
with $B$ field is that there remains
a large number of light states in an appropriate zero-slope limit
in the spectrum of $p$-$p^{\prime}$ open string \cite{SW}\cite{CIMM1}.
These states are generated by multiplying the
``almost-zero-modes".
It is natural to expect that this tower of light states should play a role
in the description of  the D$p$-brane within D$p^{\prime}$-brane
as noncommutative solitons 
both in low energy field theory and in string theory.
In order to understand the roles played by
such a large number of light states,
in this paper we compute  three point scattering amplitudes
which consist of one vertex operator of the $p^{\prime}$-$p^{\prime}$
open string
and two vertex operators of the $p$-$p^{\prime}$ open string
corresponding to the states excited by some almost-zero-modes
from the ground state.
{}From the momentum dependent multiplicative factors
in these amplitudes, we would like to read  information
as to the D$p$-brane on the noncommutative D$p^{\prime}$-brane
worldvolume.

This paper is organized as follows.
In the next section, we briefly review the worldsheet properties
of the $p$-$p^{\prime}$ open string.
In section \ref{sec:3pt}, we examine the vertex operators
of the $p$-$p^{\prime}$ open string
corresponding to the states excited by the almost-zero-modes
and calculate the three point scattering amplitudes
among two such vertex operators of the $p$-$p^{\prime}$ open string
and one vertex operator of the $p^{\prime}$-$p^{\prime}$ open string.
In this section we concentrate on bosonic string theory.
For the $p^{\prime}$-$p^{\prime}$ string vertex operators,
we prepare the tachyon vertex operator in section \ref{sec:tachyon}
and the noncommutative gauge field vertex operator
in section \ref{sec:vector}.
We find that Weyl transforms of  projection
operators emerge as multiplicative factors in the scattering
amplitudes.
In section \ref{sec:super}, we extend the analysis into
superstring theory.
The final section is devoted to summary and discussions.
In appendix \ref{sec:Weylcorr}, we present formulae of
the Weyl ordering prescription.
In appendices \ref{sec:a3tach}, \ref{sec:a3gauge} and \ref{sec:scal},
some details of the calculation of the amplitudes are given.

\section{Two Point Functions and Renormal Ordering}           
In order to fix the notation, in this section
we review some basic properties of the $p$-$p^{\prime}$ system
with a constant $B$ field.
In this paper we consider the situation where
$p<p^{\prime}$ and both of $p$ and $p^{\prime}$ are
even integers.

The setup of the system is the same as that in \cite{CIMM2},
i.e.\ a D$p$-brane extends in the $x^{0},x^{1},\ldots x^{p}$-directions
and a D$p^{\prime}$-brane extends in the
$x^{0},x^{1},\ldots x^{p^{\prime}}$-directions.
The space-time is flat with the metric
\begin{equation}
 g_{\hat{\mu}\hat{\nu}} = \left(
\begin{array}{ccccccc}
 \multicolumn{1}{c|}{-1}& & & & & & \\ \cline{1-4}
 \multicolumn{1}{c|}{ }&\multicolumn{3}{c|}{ }& & &   \\
 \multicolumn{1}{c|}{ }& &{g_{kl}}&\multicolumn{1}{c|}{}& & &  \\
 \multicolumn{1}{c|}{ }& &{ }&\multicolumn{1}{c|}{}& & & \\ \cline{2-7}
     & & &\multicolumn{1}{c|}{ } &1 & & \\
     & & &\multicolumn{1}{c|}{ } & & \ddots & \\
     & & &\multicolumn{1}{c|}{ } & & & 1 
\end{array}
\right)~,
\qquad\quad
g_{kl}=\varepsilon \delta_{kl}\quad(k,l=1,\ldots,p^{\prime})~.
\end{equation}
We bring $B_{\hat{\mu}\hat{\nu}}$ into a canonical form
\begin{equation}
B_{\hat{\mu}\hat{\nu}}=\frac{\varepsilon}{2\pi\alpha^{\prime}}\left(
 \begin{array}{ccccccccc}
   \multicolumn{1}{c|}{0}& & & & & & & & \\ \cline{1-6}
   \multicolumn{1}{c|}{}&0&b_{1}& & &\multicolumn{1}{c|}{ }& & & \\
   \multicolumn{1}{c|}{}&-b_{1}&0& & &\multicolumn{1}{c|}{ }& & &\\
   \multicolumn{1}{c|}{ }& & &\ddots& &\multicolumn{1}{c|}{ }& & &\\
   \multicolumn{1}{c|}{ }& & & &0&
        \multicolumn{1}{c|}{b_{p^{\prime}/2}}& & & \\
    \multicolumn{1}{c|}{ }& & & &-b_{p^{\prime}/2}&
        \multicolumn{1}{c|}{0}& & & \\ \cline{2-9}
    & & & & & \multicolumn{1}{c|}{ }&0& & \\
    & & & & & \multicolumn{1}{c|}{ }& &\ddots & \\
    & & & & & \multicolumn{1}{c|}{ }& & &0 \\
 \end{array}\right)~.
 \label{eq:B}
\end{equation}
In what follows, we will write the string supercoordinates along
the D$p^{\prime}$-brane as
$\mathbb{X}^{M}(\mathbf{z},\overline{\mathbf{z}})
=(\mathbb{X}^{\mu}(\mathbf{z},\overline{\mathbf{z}}),
  \mathbb{X}^{m}(\mathbf{z},\overline{\mathbf{z}}))$
$(M=0,1,\ldots,p^{\prime})$,
where $\mu(=0,1,\ldots,p)$ and $m(=p+1,\ldots,p^{\prime})$
denote the directions parallel and perpendicular to the
D$p$-brane respectively.
In terms of component fields the superfields
$\mathbb{X}^{M}(\mathbf{z},\overline{\mathbf{z}})$
are expressed as
\begin{equation}
\mathbb{X}^{M}(\mathbf{z},\overline{\mathbf{z}})
=\sqrt{\frac{2}{\alpha^{\prime}}}X^{M}(z,\overline{z})
+i\theta\psi^{M}(z)
+i\overline{\theta}\widetilde{\psi}^{M}(\overline{z})~.
\end{equation}

The string coordinates
$\mathbb{X}^{M}(\mathbf{z},\overline{\mathbf{z}})$ of
the $p^{\prime}$-$p^{\prime}$
open string and $\mathbb{X}^{\mu}(\mathbf{z},\overline{\mathbf{z}})$
of the $p$-$p^{\prime}$ open string
obey the mixed boundary conditions at the both ends.
These coordinates are expanded in an integral power
series of $z$ and $\overline{z}$ \cite{CH}\cite{AJ}\cite{CIMM2}.
The two-point functions of these string coordinates
are given in
\cite{FT}\cite{CLNY}\cite{ACNY}\cite{Garousi}\cite{SW}.
When computed on the negative real axis,
this becomes \cite{SW}\cite{CIMM2}
\begin{eqnarray}
&&\hspace{-1em}
  \mathbb{G}^{\mu\nu}
   (-e^{\tau_{1}},\theta_{1}|-e^{\tau_{2}},\theta_{2}) \equiv
 \left.\mathbb{G}^{\mu\nu}(\mathbf{z}_{1},\overline{\mathbf{z}}_{1}
                 |\mathbf{z}_{2},\overline{\mathbf{z}}_{2})
 \rule{0em}{2.5ex}
 \right|_{ \scriptstyle z_{a}=e^{\tau_{a}+i\pi} \atop \scriptstyle
           \theta_{a}=\overline{\theta}_{a}}
\equiv \left.
    \langle 0| \mathcal{R}\mathbb{X}^{\mu}
      (\mathbf{z}_{1},\overline{\mathbf{z}}_{1})
    \mathbb{X}^{\nu}(\mathbf{z}_{2},\overline{\mathbf{z}}_{2})
    |0\rangle      \rule{0em}{2.5ex}
  \right|_{\scriptstyle z_{a}=e^{\tau_{a}+i\pi} \atop
           \scriptstyle \theta_{a}=\overline{\theta}_{a}}
  \nonumber\\
&& \hspace{4em}
  =-2G^{\mu\nu}\ln(e^{\tau^{(1)}}-e^{\tau^{(2)}}
                +\theta^{(1)}\theta^{(2)})^{2}
   -\frac{i}{\alpha^{\prime}}\theta^{\mu\nu}
   \epsilon(\tau^{(1)}-\tau^{(2)})~,
  \label{eq:sw2pt-b}
\end{eqnarray}
where $\mathcal{R}$ stands for the radial ordering,
$G^{\mu\nu}$ and $\theta^{\mu\nu}$ are the inverse
of the open string metric $G_{\mu\nu}$ and 
the noncommutativity parameter
defined in \cite{SW} respectively,
and $\epsilon(x)$ denotes the sign function.

The string coordinates
$\mathbb{X}^{m}(\mathbf{z},\overline{\mathbf{z}})$
of $p$-$p^{\prime}$ open string obey the Dirichlet
boundary condition at the $\sigma=0$ end attaching
on the D$p$-brane and obey the mixed boundary
condition at the $\sigma=\pi$ end attaching
on the D$p^{\prime}$-brane.
These coordinates are expanded in a non-integral power series
of $z$ and $\overline{z}$ \cite{SW}\cite{CIMM2}.
In the following we will concentrate on the case where $p^{\prime}=p+2$.
Since we have brought the background $B_{MN}$ into the
canonical form (\ref{eq:B}),
we can readily generalize the following analysis into more
generic $p^{\prime}$ cases.
We complexify
$\mathbb{X}^{m}(\mathbf{z},\overline{\mathbf{z}})
  =(\mathbb{X}^{p+1}(\mathbf{z},\overline{\mathbf{z}}),
    \mathbb{X}^{p+2}(\mathbf{z},\overline{\mathbf{z}}))$
as
\begin{eqnarray}
&&\mathbb{Z}(\mathbf{z},\overline{\mathbf{z}})
   \equiv \mathbb{X}^{p+1}(\mathbf{z},\overline{\mathbf{z}})
   +i\mathbb{X}^{p+2}(\mathbf{z},\overline{\mathbf{z}})
 \equiv
  \sqrt{\frac{2}{\alpha^{\prime}}}Z(z,\overline{z})
  +i\theta \Psi(z)+i\overline{\theta}\widetilde{\Psi}(\overline{z})~,
  \nonumber\\
&&
\overline{\mathbb{Z}}(\mathbf{z},\overline{\mathbf{z}}) \equiv
  \mathbb{X}^{p+1}(\mathbf{z},\overline{\mathbf{z}})
    -i\mathbb{X}^{p+2}(\mathbf{z},\overline{\mathbf{z}})
   \equiv\sqrt{\frac{2}{\alpha^{\prime}}}
     \overline{Z}(z,\overline{z})
      +i\theta\overline{\Psi}(z)
      +i\overline{\theta}\widetilde{\overline{\Psi}}(\overline{z})~.
\end{eqnarray}
The mode expansions of these complex fields are
\begin{eqnarray}
&&\hspace{2em}
 Z(z,\overline{z})=i\sqrt{\frac{\alpha^{\prime}}{2}}
  \sum_{n\in \mathbb{Z}}\frac{\alpha_{n-\nu}}{n-\nu}
    \left(z^{-(n-\nu)}-\overline{z}^{-(n-\nu)}\right)~,\nonumber\\
&&\hspace{2em}
\overline{Z}(z,\overline{z})=i\sqrt{\frac{\alpha^{\prime}}{2}}
   \sum_{m\in\mathbb{Z}}\frac{\overline{\alpha}_{m+\nu}}{m+\nu}
    \left(z^{-(m+\nu)}-\overline{z}^{-(m+\nu)}\right)~, \nonumber\\
&&\Psi(z)=\sum_{r\in \mathbb{Z}+1/2}
   b_{r-\nu}z^{-(r-\nu)-\frac{1}{2}}~,\quad
   \widetilde{\Psi}(\overline{z})=-\sum_{r\in\mathbb{Z}+1/2}
   b_{r-\nu}\overline{z}^{-(r-\nu)-\frac{1}{2}}~,\nonumber\\
&&\overline{\Psi}(z)=\sum_{s\in\mathbb{Z}+1/2}
   \overline{b}_{s+\nu}z^{-(s+\nu)-\frac{1}{2}}~,\quad
   \widetilde{\overline{\Psi}}(\overline{z})
   =-\sum_{s\in\mathbb{Z}+1/2}\overline{b}_{s+\nu}
   \overline{z}^{-(s+\nu)-\frac{1}{2}}~,
 \label{eq:Zexp}
\end{eqnarray}
where $\nu$ is defined \cite{SW} by
\begin{equation}
 e^{2\pi i \nu}=-\frac{1+ib_{(p+2)/2}}{1-ib_{(p+2)/2}}~,\quad 0<\nu<1~.
\end{equation}
The modes satisfy the commutation relations
\begin{equation}
  [\alpha_{n-\nu},\overline{\alpha}_{m+\nu}]
  =\frac{2}{\varepsilon}(n-\nu)\delta_{n+m}~,
 \quad \{b_{r-\nu},\overline{b}_{s+\nu}\}
 =\frac{2}{\varepsilon}\delta_{r+s}~.
  \label{eq:com}
\end{equation}
The fact that the string coordinates $Z(z,\overline{z})$
and $\overline{Z}(z,\overline{z})$ are expanded by
a non-integral power series of $z$ and $\overline{z}$
implies that a twist field $\sigma^{+}(z,\overline{z})$
and an anti-twist field $\sigma^{-}(z,\overline{z})$,
both of which are mutually non-local with respect to $Z$ and $\overline{Z}$,
locate at the origin and at the infinity on the complex plane respectively
and generate a branch-cut between themselves.
{}For the fermionic coordinates the situation is the same.
The mutually non-local fields
in the fermionic sector are a spin field $s^{+}(z,\overline{z})$
and an anti-spin field $s^{-}(z,\overline{z})$.
The (anti-) twist and the (anti-) spin fields
 serve as boundary condition changing
operators \cite{AH}\cite{GNS}.
The fields $\sigma^{+}$ and $s^{+}$ create the incoming oscillator
vacuum $|\sigma,s\rangle$ and the fields $\sigma^{-}$ and $s^{-}$ create
the outgoing vacuum $\langle \sigma,s|$ \cite{CIMM2}.
The two-point function of $\mathbb{Z}$ and $\overline{\mathbb{Z}}$
evaluated on these vacua are obtained in \cite{CIMM2}.
On the negative real axis,
i.e.\ on the D$(p+2)$-brane worldvolume, it takes the form
\cite{CIMM2} of
\begin{eqnarray}
&&\hspace{-1em}\mbox{\boldmath$\mathcal{G}$}^{Z\overline{Z}}
        (-e^{\tau_{1}},\theta_{1}|-e^{\tau_{2}},\theta_{2})
\equiv \left.\mbox{\boldmath$\mathcal{G}$}^{Z\overline{Z}}
        (\mathbf{z}_{1},\overline{\mathbf{z}}_{1}
         |\mathbf{z}_{2},\overline{\mathbf{z}}_{2})\rule{0em}{2.5ex}
\right|_{\scriptstyle z_{a}=e^{\tau_{a}+i\pi}
         \atop \scriptstyle \theta_{a}=\overline{\theta}_{a}}
\equiv \left. \langle \sigma,s | \mathcal{R}
   \mathbb{Z}(\mathbf{z}_{1},\overline{\mathbf{z}}_{1})
  \overline{\mathbb{Z}}(\mathbf{z}_{2},\overline{\mathbf{z}}_{2})
  |\sigma,s\rangle \rule{0em}{2.5ex}
  \right|_{\scriptstyle z_{a}=e^{\tau_{a}+i\pi}
           \atop \scriptstyle \theta_{a}=\overline{\theta}_{a}}
\nonumber\\
&&  \hspace{2em}
=\frac{8}{\varepsilon(1+b^{2})}
\left[\Theta(\tau_{1}-\tau_{2})
 \mathcal{F}\left(1-\nu;\frac{e^{\tau_{2}}-\theta_{1}\theta_{2}}
                             {e^{\tau_{1}}}\right)
  +\Theta(\tau_{2}-\tau_{1})
 \mathcal{F}\left(\nu;\frac{e^{\tau_{1}}}
                           {e^{\tau_{2}}-\theta_{1}\theta_{2}}\right)
 \right]~,
\end{eqnarray}
where $\Theta(x)$ is the step function and $\mathcal{F}(\nu;z)$
is defined by using the hypergeometric function
$F(a,b;c;z)$ as
\begin{equation}
 \mathcal{F}(\nu;z)=\frac{z^{\nu}}{\nu}F(1,\nu;1+\nu;z)
  =\sum_{n=0}^{\infty}\frac{1}{n+\nu}z^{n+\nu}~.
\end{equation}

In this way, this system has two types of vacuum
for the string coordinates $\mathbb{Z}$ and
 $\overline{\mathbb{Z}}$;
the one is the $SL(2,\mathbb{R})$ invariant vacuum $|0\rangle$ and
the other is the oscillator vacuum $|\sigma,s\rangle$.
This implies that we can define two types of normal ordering
associated with the vacua $|0\rangle$ and $|\sigma,s \rangle$
respectively.
We denote the normal ordering associated with $|0\rangle$
by $:\ \,:$ and that associated with $|\sigma,s\rangle$
by $\maru~\;\maru$. 
The relation between these normal orderings is described by
the renormal ordering formula~\cite{CIMM2}:
\begin{eqnarray}
&&:\mbox{\boldmath$\mathcal{O}$}:\;
=\exp\left(\int \!d^{2}\mathbf{z}_{1}d^{2}\mathbf{z}_{2}
   \ \mbox{\boldmath$\mathcal{G}$}^{Z\overline{Z}}_{\mathrm{sub}}
       (\mathbf{z}_{1},\overline{\mathbf{z}}_{1}
         |\mathbf{z}_{2},\overline{\mathbf{z}}_{2})
      \frac{\delta}{\delta \mathbb{Z}
                    (\mathbf{z}_{1},\overline{\mathbf{z}}_{1})}
      \frac{\delta}{\delta \overline{\mathbb{Z}}
                    (\mathbf{z}_{2},\overline{\mathbf{z}}_{2})}
               \right)
  \maru~\mbox{\boldmath$\mathcal{O}$}~\maru~,\nonumber\\
 && \hspace{2em}
 \mbox{\boldmath$\mathcal{G}$}^{Z\overline{Z}}_{\mathrm{sub}}
   (\mathbf{z}_{1},\overline{\mathbf{z}}_{1}
      |\mathbf{z}_{2},\overline{\mathbf{z}}_{2})
    \equiv \mbox{\boldmath$\mathcal{G}$}^{Z\overline{Z}}
       (\mathbf{z}_{1},\overline{\mathbf{z}}_{1}
         |\mathbf{z}_{2},\overline{\mathbf{z}}_{2})
      -\mathbb{G}^{Z\overline{Z}}
       (\mathbf{z}_{1},\overline{\mathbf{z}}_{1}
        |\mathbf{z}_{2},\overline{\mathbf{z}}_{2})~,
 \label{eq:renormal}
\end{eqnarray}
where {\boldmath$\mathcal{O}$} is an arbitrary functional of
$\mathbb{Z}$ and $\overline{\mathbb{Z}}$
and  $\mathbb{G}^{Z\overline{Z}}(\mathbf{z}_{1},\overline{\mathbf{z}}_{1}
      |\mathbf{z}_{2},\overline{\mathbf{z}}_{2})$
is complexification of
$\mathbb{G}^{mn}(\mathbf{z}_{1},\overline{\mathbf{z}}_{1}
                  |\mathbf{z}_{2},\overline{\mathbf{z}}_{2})$:
\begin{eqnarray}
 &&\mathbb{G}^{Z\overline{Z}}(\mathbf{z}_{1},\overline{\mathbf{z}}_{1}
                             |\mathbf{z}_{2},\overline{\mathbf{z}}_{2})
 =\mathbb{G}^{p+1,p+1}(\mathbf{z}_{1},\overline{\mathbf{z}}_{1}
                       |\mathbf{z}_{2},\overline{\mathbf{z}}_{2})
  +\mathbb{G}^{p+2,p+2}(\mathbf{z}_{1},\overline{\mathbf{z}}_{1}
                      |\mathbf{z}_{2},\overline{\mathbf{z}}_{2})
 \nonumber\\
  &&\hspace{8em}
   -i\mathbb{G}^{p+1,p+2}(\mathbf{z}_{1},\overline{\mathbf{z}}_{1}
                         |\mathbf{z}_{2},\overline{\mathbf{z}}_{2})
  -i\mathbb{G}^{p+2,p+1}(\mathbf{z}_{1},\overline{\mathbf{z}}_{1}
                        |\mathbf{z}_{2},\overline{\mathbf{z}}_{2})~.
\end{eqnarray}

We will take the zero-slope limit proposed by Seiberg and Witten
\cite{SW} in which $\alpha^{\prime}$ is sent to zero with
the open string metric $G_{MN}$ and the noncommutativity parameter
$\theta^{MN}$ kept finite. This implies that
\begin{equation}
\left\{
\begin{array}{l}
\alpha^{\prime}\sim \varepsilon^{1/2}\rightarrow 0\\
g_{MN}\sim \varepsilon\rightarrow 0\\
 |b_{I}| \sim \varepsilon^{-1/2}\rightarrow \infty
  \quad (I=1,\ldots,(p+2)/2)
\end{array}
\right.~.\label{eq:zero-slope}
\end{equation}
In this limit, whether $b_{I}$ goes to $+\infty$ or $-\infty$
just depends on the convention.
In what follows, we will send $b_{(p+2)/2}$ to $+\infty$
in the zero-slope limit. Thus in the zero-slope limit we have
\begin{eqnarray}
&& \theta^{p+1\, p+2}
 \equiv -\frac{2\pi\alpha^{\prime}b_{(p+2)/2}}
               {\varepsilon \left(1+\left(b_{(p+2)/2}\right)^{2}\right)}
        <0~,\nonumber\\
&&1-\nu \simeq \frac{1}{\pi b_{(p+2)/2}} \sim \varepsilon^{1/2}
\rightarrow 0~.  \label{eq:theta}
\end{eqnarray}

\section{Three Point Amplitudes and Multiplicative Factors
in Bosonic String Theory}
\label{sec:3pt}

We will calculate three point amplitudes which consist of
two vertex operators of the $p$-$(p+2)$ open string
and one vertex operator of the $(p+2)$-$(p+2)$ open string.
As we mentioned in the introduction,
in this section we restrict ourselves to bosonic string theory.
The computations in this section will help us to
extend these investigations into the superstring case
in the next section.

The open string vertex operators should be inserted at the
boundary of the worldsheet, namely the real axis of the
complex upper half plane ($z$-plane).
We insert at $\xi\equiv \mathrm{Re}(z)=\xi^{(1)}$
and at $\xi=\xi^{(3)}(>\xi^{(1)})$
the vertex operators of the $p$-$(p+2)$ open string
which contain the anti-twist field $\sigma^{-}$ and twist field
$\sigma^{+}$ respectively.
Since the (anti-) twist field serves as the boundary condition changing
operator, the interval of the worldsheet boundary
between them, $\xi \in [\xi^{(1)},\xi^{(3)}]$, is interpreted to be
on the D$(p+2)$-brane and the remaining part
be on the D$p$-brane \cite{AH}\cite{GNS}.
Thus the location $\xi^{(2)}$ of the vertex operator
of the $(p+2)$-$(p+2)$ open string should be
$\xi^{(2)}\in[\xi^{(1)},\xi^{(3)}]$.

\subsection{Vertex operators of $p$-$(p+2)$ open string}

In the zero-slope limit (\ref{eq:zero-slope}),
there exist {\it almost-zero-modes\/} $\overline{\alpha}_{-1+\nu}$
and $\alpha_{1-\nu}$ in the bosonic sector
of the $p$-$(p+2)$ open string.
In the superstring case,
these modes generate a tower of light states
which survives
the zero slope limit~\cite{SW}\cite{CIMM1}.
This implies that in the low energy physics
D$p$-brane is perturbatively seen as a collection of
such a large number of light states.
This fact gives the possibility that
we grasp some nontrivial D-brane physics in the
perturbative analysis even after taking the zero-slope limit.

We consider the vertex operators
$\mathcal{V}^{(+)}_{\mn}(\xi;k_{\mu})$ and
$\mathcal{V}^{(-)}_{\mm}(\xi;k_{\mu})$ 
of the $p$-$(p+2)$ open string
which correspond to the states excited by almost-zero-modes from
the oscillator vacuum $|\sigma\rangle$:
\begin{eqnarray}
&&\mathcal{V}^{(+)}_{\mn}(\xi;k_{\mu})=
   :\exp \left( i k_{\mu}X^{\mu}\right) (\xi):V^{(+)}_{\mn}(\xi)~,
   \nonumber\\
&& \mathcal{V}^{(-)}_{\mm}(\xi;k_{\mu})=
   :\exp \left(i k_{\mu}X^{\mu}\right)(\xi):V^{(-)}_{\mm}(\xi)~,
\end{eqnarray}
where $V^{(+)}_{\mn}(\xi)$ and
$V^{(-)}_{\mm}(\xi)$ are the contributions from the string coordinates
$Z$ and $\overline{Z}$ defined by
\begin{eqnarray}
 &&\lim_{\xi \rightarrow 0}V^{(+)}_{\mn}(\xi) |0\rangle
  =(\overline{\alpha}_{-1+\nu})^{\mn}|\sigma\rangle
  \qquad \quad \;(\mn=0,1,2\ldots)~,
  \nonumber\\
 &&\lim_{\xi \rightarrow\infty} \xi^{2h}\langle 0| V^{(-)}_{\mm}(\xi)
 =\langle \sigma | (\alpha_{1-\nu})^{\mm}
 \qquad (\mm=0,1,2,\ldots)~.
 \label{eq:vnm}
\end{eqnarray}
Here $h$ is the conformal weight of the operator $V^{(-)}_{\mm}(\xi)$,
\begin{equation}
h=\left(\mm+\frac{\nu}{2}\right)(1-\nu)~.
\end{equation}
We note that the momentum $k_{\mu}$ carried by
$\mathcal{V}_{\mm}^{(\pm)}(\xi;k_{\mu})$ is a $(p+1)$ vector
along the D$p$-brane worldvolume.
Taking the relation
$(\alpha_{1-\nu})^{\dagger}=\overline{\alpha}_{-1+\nu}$
into account,
one finds that the definition (\ref{eq:vnm}) means that
the space-time field $\Phi_{\mm}^{(-)}$ corresponding to the vertex
operator $\mathcal{V}_{\mm}^{(-)}(\xi;k_{\mu})$ is
hermitian conjugate to the field $\Phi_{\mm}$ corresponding to
$\mathcal{V}_{\mm}^{(+)}(\xi;k_{\mu})$:
$\Phi_{\mm}^{(-)}=\Phi^{\dagger}_{\mm}$.
The physical state conditions for
$\mathcal{V}^{(+)}_{\mm}(\xi;k_{\mu})$ and
$\mathcal{V}^{(-)}_{\mm}(\xi;k_{\mu})$
require
the on-shell condition
$\alpha^{\prime}m^{2}
  \equiv -\alpha^{\prime} G^{\mu\nu}k_{\mu}k_{\nu}
  =(\mm+\frac{\nu}{2})(1-\nu)-1$
for the field $\Phi_{\mm}^{(\dagger)}$.

\subsection{Probe with tachyon field}\label{sec:tachyon}

In this subsection we examine the process probing the D$p$-brane
with the tachyon field excited from the $(p+2)$-$(p+2)$ open string.
The vertex operator of the tachyon is
\begin{equation}
\mathcal{V}_{\phi}(\xi;k_{M})
=:\exp\left( i\sum_{M=0}^{p+2}k_{M}X^{M}\right)(\xi):
 =:\exp\left( i\sum_{\mu=0}^{p}k_{\mu}X^{\mu}\right)(\xi):
 V_{\phi}(\xi;k_{m})~, \label{eq:vtachyon}
\end{equation}
where $V_{\phi}(\xi;k_{m})$ is the contribution from the fields
$Z$ and $\overline{Z}$:
\begin{eqnarray}
&&V_{\phi}(\xi;k_{m})=:\exp\left[i\left(\kappa Z
                   +\overline{\kappa}\overline{Z}\right)\right](\xi):~,
     \nonumber\\
&& \hspace{1.5em}\mbox{with}\quad
\kappa\equiv \frac{1}{2}\left(k_{p+1}-ik_{p+2}\right)~,
\quad \overline{\kappa}\equiv \frac{1}{2}\left(k_{p+1}+ik_{p+2}\right)~.
\end{eqnarray}
Note that the momenta carried by the vertex operators of
the $(p+2)$-$(p+2)$ open string are $(p+2)+1$ dimensional vectors
in the space-time
while those carried by the $p$-$(p+2)$ open string are
$p+1$ dimensional ones.
The physical state conditions for
the vertex operator $\mathcal{V}_{\phi}(\xi;k_{M})$
require that the on-shell condition
$\alpha^{\prime}m^{2}\equiv -\alpha^{\prime}k_{M}k_{N}G^{MN}=-1$
should hold.

We would like to calculate the three point disc amplitude
\begin{equation}
\mathcal{A}_{3}^{\mathrm{tachyon}}
= c \int \frac{d\xi^{(1)}d\xi^{(2)}d\xi^{(3)}}
              {d^{3}F(\xi^{(1)},\xi^{(2)},\xi^{(3)})}
  \langle 0 | \,
  \mathcal{V}^{(-)}_{\mm}\left(\xi^{(1)};k^{(1)}_{\mu}\right)\,
  \mathcal{V}_{\phi}\left(\xi^{(2)};k^{(2)}_{M}\right)\,
  \mathcal{V}^{(+)}_{\mn}\left(\xi^{(3)};k_{\mu}^{(3)}\right)\,
       |0\rangle~,       \label{eq:a3}
\end{equation}
where $c$ is an overall constant which we are not interested in,
and $d^{3}F(\xi^{(1)},\xi^{(2)},\xi^{(3)})$ denotes the volume element
of the isometry $SL(2,\mathbb{R})$ gauge group generated
by the conformal Killing vectors
on the worldsheet:
\begin{equation}
d^{3}F(\xi^{(1)},\xi^{(2)},\xi^{(3)})
 =\frac{d\xi^{(1)}d\xi^{(2)}d\xi^{(3)}}
       {(\xi^{(1)}-\xi^{(2)})(\xi^{(2)}-\xi^{(3)})(\xi^{(3)}-\xi^{(1)})}~.
\end{equation}
This amplitude measures the process where the field $\Phi_{\mn}$
propagating  along the D$p$-brane with momentum $k^{(3)}_{\mu}$
prepared at the initial state is scattered by the tachyon on the
D$(p+2)$-brane and then at the final state the field $\Phi_{\mm}$
is observed moving along the D$p$-brane with momentum
$-k^{(2)}_{\mu}$.

Through the computations presented in appendix \ref{sec:a3tach},
we obtain
\begin{eqnarray}
&& \mathcal{A}_{3}^{\mathrm{tachyon}}
 =c  \prod_{\mu=0}^{p}\delta(k^{(1)}_{\mu}+k^{(2)}_{\mu}+k^{(3)}_{\mu})
 \exp \left[\sum_{1\leq c<d \leq 3}\frac{i}{2}\theta^{\mu\nu}
               k^{(c)}_{\mu}k^{(d)}_{\nu}
               \epsilon(\tau^{(c)}-\tau^{(d)})\right]
     \nonumber\\
 &&\hspace{5em}
   \times \sqrt{\mm!\,\mn !} \, 
   \left(\frac{2(1-\nu)}{\varepsilon}\right)^{\frac{\mm+\mn}{2}}
     \mathcal{I}_{\mn\mm} \left(b_{(p+2)/2},k^{(2)}_{m}\right)~,
 \label{eq:a3-f}
\end{eqnarray}
where the momentum dependent factor
$\mathcal{I}_{\mn\mm} (b_{(p+2)/2},k^{(2)}_{m})$ is
defined as
\begin{equation}
\mathcal{I}_{\mn\mm} \equiv\left\{
  \begin{array}{ll}
  \displaystyle \sqrt{\frac{\mn!}{\mm !}}
  \left( i\overline{\kappa}^{(2)}\sqrt{2\Xi}\right)^{\mm-\mn}
  \exp \mathcal{C}(\nu;k^{(2)}_{m})
  L_{\mn}^{(\mm-\mn)}(2\Xi\kappa^{(2)}\overline{\kappa}^{(2)})
  & (\mm \geq \mn)  \\[1.5ex]
  \displaystyle \sqrt{\frac{\mm !}{\mn !}}
  \left(i\kappa^{(2)}\sqrt{2\Xi}\right)^{\mn-\mm}
  \exp\mathcal{C}(\nu;k^{(2)}_{m})
   L_{\mm}^{(\mn-\mm)}(2\Xi\kappa^{(2)}\overline{\kappa}^{(2)})
   & (\mm<\mn)
  \end{array}
\right. ,  \label{eq:ftach}
\end{equation}
with
\begin{equation}
\Xi \equiv \frac{2\alpha^{\prime}}
            {\varepsilon (1-\nu) \left(1+(b_{(p+2)/2})^{2}\right)}~.
\end{equation}
In eq.(\ref{eq:ftach}),
$L^{(\alpha)}_{\mn}(x)$ denotes the Laguerre polynomial
\begin{equation}
L^{(\alpha)}_{\mn}(x)
  =\frac{x^{-\alpha}e^{x}}{\mn!}
     \frac{d^{\mn}}{dx^{\mn}}\left(e^{-x}x^{\mn+\alpha}\right)
  =\sum_{l=0}^{\mn}(-1)^{l}
    \left( \begin{array}{c}
                \mn+\alpha \\
                \mn-l
     \end{array} \right)\frac{x^{l}}{l!}~,
   \label{eq:Laguerre}
\end{equation}
$\mathcal{C}(\nu;k_{m})$ is defined as
\begin{eqnarray}
\label{eq:cfactor}
&&\mathcal{C}(\nu;k_{m})=2\alpha^{\prime}
  \left\{ \gamma+\frac{1}{2}\left(\psi(\nu)+\psi(1-\nu)\right)\right\}
  G^{Z\overline{Z}}\kappa\overline{\kappa}~, \\
&&\hspace{3em} \gamma : \mbox{ Euler constant}~,\quad
               \psi(\nu) : \mbox{ digamma function}~,\nonumber
\end{eqnarray}
and $G^{Z\overline{Z}}$ is the open string metric
in the complex notation in the $x^{p+1},x^{p+2}$ directions:
\begin{equation}
G^{Z\overline{Z}}=G^{p+1\, p+1}+G^{p+2\, p+2}-2iG^{p+1\,p+2}
=\frac{2}{\varepsilon\left(1+\left(b_{(p+2)/2}\right)^{2}\right)}~.
\end{equation}

Here we make some comments on the momentum dependent
multiplicative factor $\mathcal{I}_{\mn\mm}$.
The gaussian damping factor
$\exp\mathcal{C}(\nu;k^{(2)}_{m})(=\mathcal{I}_{00})$
included in all of the multiplicative factors $\mathcal{I}_{\mn\mm}$
comes from the scattering process
in which the almost-zero modes
do not participate.
This factor takes the same form as what is obtained in the superstring
case \cite{CIMM2}.
The every other piece in $\mathcal{I}_{\mn\mm}$
is the contribution from the almost-zero-modes.
This originates from eq.(\ref{eq:mVn}).

Let us take the zero-slope limit (\ref{eq:zero-slope}).
In this limit,
$\mathcal{C}(\nu;k^{(2)}_{m})
  \rightarrow -|\theta^{p+1\,p+2}|\kappa^{(2)}\overline{\kappa}^{(2)}$
\cite{CIMM2}\cite{CIMM3} and
$\Xi \rightarrow |\theta^{p+1\,p+2}|$.
This tells us that
\begin{equation}
 \mathcal{I}_{\mn\mm}
 \rightarrow 
\left\{
    \begin{array}{ll}
    \displaystyle
    \sqrt{\frac{\mn !}{\mm !}}
      \left(i\sqrt{2|\theta^{p+1\,p+2}| }
                   \overline{\kappa}^{(2)} \right)^{\mm-\mn}
        e^{-|\theta^{p+1\,p+2}|\kappa^{(2)}\overline{\kappa}^{(2)}}
        L^{(\mm-\mn)}_{\mn}
         \left(2|\theta^{p+1\,p+2}|\kappa^{(2)}\overline{\kappa}^{(2)}
         \right) & \ (\mm\geq \mn)\\[1.5ex]
     \displaystyle
     \sqrt{\frac{\mm !}{\mn !}}
      \left(i\sqrt{2|\theta^{p+1\,p+2}|}
                     \kappa^{(2)} \right)^{\mn-\mm}
        e^{-|\theta^{p+1\,p+2}|\kappa^{(2)}\overline{\kappa}^{(2)}}
        L^{(\mn-\mm)}_{\mm}\left(
         2|\theta^{p+1\,p+2}| \kappa^{(2)}\overline{\kappa}^{(2)}
         \right) &\ (\mm < \mn)
    \end{array}
  \right.~.
\end{equation}
This is precisely the Weyl transform
$\widetilde{f}_{\mn\mm}(k_{m})$ of the operator
$|\mn\rangle\langle \mm|$ on the noncommutative
$\mathbb{R}^{2}$ space obtained in eq.(\ref{eq:weylnm}).
Consequently the momentum dependent multiplicative factor
$\mathcal{I}_{\mn\mm}$ which we read off from the amplitude for the process
probing the D$p$-brane with the tachyon field
becomes the Weyl transform of the operator $|\mn\rangle\langle\mm|$
in the zero-slope limit.
In the case of $\mm=\mn$ in particular,
this becomes the projection operator
$|\mm\rangle\langle \mm|$.
This is what we wanted.


\subsection{Probe with gauge field}\label{sec:vector}

In this subsection we would like to probe the D$p$-brane
with the gauge field excited from the $p^{\prime}$-$p^{\prime}$
open string. The vertex operator of the gauge field is
\begin{equation}
\mathcal{V}_{A}(\xi;k_{M};\zeta_{M})
 = i : \zeta_{M}(k)\dot{X}^{M}(\xi)\,\exp (ik_{N}X^{N}) (\xi):~,
\end{equation}
where $\dot{X}^{M}(\xi)\equiv (\partial+\overline{\partial})
         \left.X^{M}(z,\overline{z})\right|_{z=\overline{z}=\xi}$.
The physical state conditions for this vertex operator require
the on-shell condition
$\alpha^{\prime}m^{2}\equiv -\alpha^{\prime}
  G^{MN}k_{M}k_{N}=0$
and the polarization condition
$G^{MN}k_{M}\zeta_{N}=0$.

Let us compute the three point disc amplitude
\begin{equation}
\mathcal{A}^{\mathrm{gauge}}_{3}= \tilde{c}
  \int \frac{d\xi^{(1)} d\xi^{(2)} d\xi^{(3)}}
            {d^{3}F(\xi^{(1)},\xi^{(2)},\xi^{(3)})}
   \langle 0 | \mathcal{V}^{(-)}_{\mm}(\xi^{(1)};k^{(1)}_{\mu})
     \,\mathcal{V}_{A}(\xi^{(2)};k^{(2)}_{M};\zeta_{M})
     \,\mathcal{V}^{(+)}_{\mn}(\xi^{(3)};k^{(3)}_{\mu})
   | 0 \rangle~.\label{eq:a3vec}
\end{equation}
Following the calculation given in appendix \ref{sec:a3gauge},
we find that
\begin{eqnarray}
&& \mathcal{A}^{\mathrm{gauge}}_{3}
= \tilde{c}\prod_{\mu=0}^{p}
   \delta(k^{(1)}_{\mu}+k^{(2)}_{\mu}+k^{(3)}_{\mu})
  \prod_{1\leq c<d \leq 3}\exp
   \left[\frac{i}{2}\theta^{\mu\lambda}k^{(c)}_{\mu}k^{(d)}_{\lambda}
         \epsilon(\tau^{(c)}-\tau^{(d)}) \right]
        \nonumber\\
&&\hspace{5em} \times
  \left(\frac{2(1-\nu)}{\varepsilon}\right)^{\frac{\mm+\mn}{2}}
   \sqrt{\mm!\,\mn!} \; \alpha^{\prime}\;\,
   \mathcal{K}_{\mn\mm}~,
  \label{eq:a3vec-f}
\end{eqnarray}
where
$\mathcal{K}_{\mn\mm}$
is defined for the $\mm\geq \mn$ case as
\begin{eqnarray}
  &&\hspace{-2em}\mathcal{K}_{\mn\leq \mm}=
  \sqrt{\frac{\mn!}{\mm!}}
  \exp\mathcal{C}(\nu;k^{(2)}_{m})
\Bigg[
  \left\{ -(k^{(3)}-k^{(1)})\pdot\zeta
          +(2\nu-1) G^{Z\overline{Z}}
            (\kappa^{(2)}\overline{e}-\overline{\kappa}^{(2)}e)
      \right\}\nonumber\\
 &&\hspace{15em}\times
    \left(i\sqrt{2\Xi}\overline{\kappa}^{(2)}\right)^{\mm-\mn}
    L^{(\mm-\mn)}_{\mn}\left(2\Xi \kappa^{(2)}\overline{\kappa}^{(2)}
                       \right)
   \nonumber\\
 && \hspace{6em}+2G^{Z\overline{Z}}
    (\kappa^{(2)}\overline{e}-\overline{\kappa}^{(2)}e)
    \left(i\sqrt{2\Xi}\overline{\kappa}^{(2)}\right)^{\mm-\mn}
    L^{(\mm-\mn+1)}_{\mn-1}\left(
         2\Xi\kappa^{(2)}\overline{\kappa}^{(2)}\right)\nonumber\\  
&&   \hspace{6em}-i(\mm-\mn)\frac{1-\nu}{\alpha^{\prime}}\overline{e}
   \sqrt{2\Xi}  
   \left(i\sqrt{2\Xi}\overline{\kappa}^{(2)}
   \right)^{\mm-\mn-1}
     L^{(\mm-\mn)}_{\mn}
   \left(2\Xi\kappa^{(2)}\overline{\kappa}^{(2)}\right)
   \Bigg]~,~~~ \label{eq:kln<m}
\end{eqnarray}
and for the $\mm<\mn$ case as
\begin{eqnarray}
&& \hspace{-2em} \mathcal{K}_{\mn>\mm}=
    \sqrt{\frac{\mm!}{\mn!}}\exp\mathcal{C}(\nu;k^{(2)}_{m})\Bigg[
 \left\{-(k^{(3)}-k^{(1)})\pdot\zeta
 +(2\nu-1)G^{Z\overline{Z}}
       (\kappa^{(2)}\overline{e}-\overline{\kappa}^{(2)}e)
  \right\}  \nonumber\\
&& \hspace{15em} \times 
   \left( i \sqrt{2\Xi}\kappa^{(2)} \right)^{\mn-\mm}
   L^{(\mn-\mm)}_{\mm}
     \left(2\Xi \kappa^{(2)}\overline{\kappa}^{(2)}\right)
  \nonumber\\
&& \hspace{6em} +2G^{Z\overline{Z}}
  (\kappa^{(2)}\overline{e}-\overline{\kappa}^{(2)}e)
  \left(i\sqrt{2\Xi}\kappa^{(2)}\right)^{\mn-\mm}
  L^{(\mn-\mm+1)}_{\mm-1}
    \left(2\Xi \kappa^{(2)}\overline{\kappa}^{(2)}\right)\nonumber\\
&& \hspace{6em} +i(\mn-\mm) \frac{(1-\nu)}{\alpha^{\prime}}e
    \sqrt{2\Xi} 
     \left(i\sqrt{2\Xi}\kappa^{(2)}\right)^{\mn-\mm-1}
    L^{(\mn-\mm)}_{\mm}
      \left(2\Xi \kappa^{(2)}\overline{\kappa}^{(2)}\right)\Bigg]~.~~~
   \label{eq:kln>m}
\end{eqnarray}
Here the symbol ``$\pdot$" denotes the inner product
of the $p+1$ dimensional vectors along the D$p$-brane
worldvolume with respect to the open string metric:
$A\pdot B=G^{\mu\lambda}A_{\mu}B_{\lambda}$,
and $e$ and $\overline{e}$ are the polarization tensors
$\zeta_{p+1}(k_{M})$ and $\zeta_{p+2}(k_{M})$ in the complex
notation defined as
\begin{equation}
 e(k_{M})\equiv \frac{1}{2}
       \left(\zeta_{p+1}(k_{M})-i\zeta_{p+2}(k_{M})\right)~,
 \quad \overline{e}(k_{M}) \equiv \frac{1}{2}
       \left(\zeta_{p+1}(k_{M})+i\zeta_{p+2}(k_{M})\right)~.
  \label{eq:eebar}
\end{equation}

By using the formula for the Laguerre polynomial
\begin{equation}
 L_{\mn}^{(\beta)}(x) = \sum_{\mathrm{k}=0}^{\mn}
  \frac{\Gamma(\mn-\mathrm{k}+\beta-\alpha)}
        {\Gamma(\mn-\mathrm{k}+1)\Gamma(\beta-\alpha)}
    L^{(\alpha)}_{\mathrm{k}} (x)~, \label{eq:lageq}
\end{equation}
we can recast the factor
$\mathcal{K}_{\mn\mm}$ as
\begin{eqnarray}
&&\mathcal{K}_{\mn\leq\mm}\equiv
\left\{-(k^{(3)}-k^{(1)})\pdot \zeta
       +(2\nu-1)G^{Z\overline{Z}}
        (\kappa^{(2)}\overline{e}-\overline{\kappa}^{(2)}e)
    \right\} \mathcal{I}_{\mn\mm}
 \nonumber\\
&& \hspace{3.5em}+2G^{Z\overline{Z}}
  (\kappa^{(2)}\overline{e}-\overline{\kappa}^{(2)}e)
  \sqrt{\frac{\mn!}{\mm!}}
  \,\sum_{\mathrm{k}=0}^{\mn -1}
    \sqrt{\frac{(\mm-\mn+\mathrm{k})!}{\mathrm{k}!}}
    \,\mathcal{I}_{\mathrm{k}\;\mm-\mn+\mathrm{k}}
   \nonumber\\
&& \hspace{3.5em}  -i(\mm-\mn)\frac{1-\nu}{\alpha^{\prime}}
   \overline{e}\sqrt{2\Xi}\sqrt{\frac{\mn!}{\mm!}}
   \,\sum_{\mathrm{k}=0}^{\mn}
     \sqrt{\frac{(\mm-\mn-1+\mathrm{k})!}{\mathrm{k}!}}
     \,\mathcal{I}_{\mathrm{k}\;\mm-\mn-1+\mathrm{k}}~,
   \label{eq:kn<m}
\end{eqnarray}
and
\begin{eqnarray}
&&\mathcal{K}_{\mn>\mm} \equiv \left\{
 -(k^{(3)}-k^{(1)})\pdot \zeta
 +(2\nu-1)G^{Z\overline{Z}}
 (\kappa^{(2)}\overline{e}-\overline{\kappa}^{(2)}e)\right\}
 \mathcal{I}_{\mn\mm}  \nonumber\\
&& \hspace{3.5em} +2G^{Z\overline{Z}}
   (\kappa^{(2)}\overline{e}-\overline{\kappa}^{(2)}e)
   \sqrt{ \frac{\mm!}{\mn!} }
   \, \sum_{\mathrm{k}=0}^{\mm-1}
      \sqrt{\frac{(\mn-\mm+\mathrm{k})!}{\mathrm{k}!}}
   \,\mathcal{I}_{\mn-\mm+\mathrm{k}\;\mathrm{k}}
 \nonumber\\
&&\hspace{3.5em}
 -i (\mn-\mm)\frac{1-\nu}{\alpha^{\prime}}e
   \sqrt{2\Xi}\sqrt{\frac{\mm!}{\mn!}}
   \,\sum_{\mathrm{k}=0}^{\mm}
   \sqrt{\frac{(\mn-\mm-1+\mathrm{k})!}{\mathrm{k}!}}
   \,\mathcal{I}_{\mn-\mm-1+\mathrm{k}\;\mathrm{k}}~,
  \label{eq:kn>m}
\end{eqnarray}
where $\mathcal{I}_{\mn\mm}$ is the momentum dependent factor
defined in eq.(\ref{eq:ftach})
and $\sum_{\mathrm{k}=0}^{\mn-1}$ means zero if $\mn=0$.

The first line in each of eqs.(\ref{eq:kn<m}) and (\ref{eq:kn>m})
has a similar feature to the scattering amplitude
$\mathcal{A}^{\mathrm{tachyon}}_{3}$
in the sense that it can be obtained from
the amplitude of the $\mm=\mn=0$ case
by replacing the gaussian damping factor
$ \exp \mathcal{C}(\nu;k^{(2)}_{m})$ with
$\mathcal{I}_{\mn\mm}(b_{(p+2)/2},k^{(2)}_{m})$.
The second and the third lines
in each of eqs.(\ref{eq:kn<m}) and (\ref{eq:kn>m})
originate in the interactions between the almost-zero-modes
and the gauge field on D$(p+2)$-brane worldvolume.

In the case of $\mm=\mn$, $\mathcal{K}_{\mm\mm}$ has a simple form:
\begin{eqnarray}
&&\mathcal{K}_{\mm\mm}=
\left\{ -(k^{(3)}-k^{(1)}) \pdot \zeta
 +(2\nu-1)G^{Z\overline{Z}}
 (\kappa^{(2)}\overline{e}-\overline{\kappa}^{(2)}e)
\right\}
 \, \mathcal{I}_{\mm\mm}\nonumber\\
&&\hspace{3em}
 +2G^{Z\overline{Z}}(\kappa^{(2)}\overline{e}-\overline{\kappa}^{(2)}e)
   \sum_{\mathrm{k}=0}^{\mm-1}\mathcal{I}_{\mathrm{k k}}~.
\end{eqnarray}
In the zero slope limit,
the multiplicative factors
$\mathcal{I}_{\mm\mm}$ and
$\sum_{\mathrm{k}=0}^{\mm-1}\mathcal{I}_{\mathrm{k k}}$
in this equation become projection operators
$|\mm\rangle\langle \mm|$ and
$\sum_{\mathrm{k}=0}^{\mm-1}|\mathrm{k}\rangle\langle \mathrm{k}|$
respectively.
These are what we desired to obtain.

\section{Three Point Amplitudes and Multiplicative Factors
  in Superstring Theory}\label{sec:super}
In this section we extend the analyses in the last section
into the superstring case.
We study the process in which the D$p$-brane is
probed with the noncommutative gauge field on the D$(p+2)$-brane
worldvolume.

\subsection{Vertex operators of light states of $p$-$(p+2)$
open string}

We would like to obtain  the physical state
of the $p$-$(p+2)$ open string
which is excited by almost-zero-modes $\overline{\alpha}_{-1+\nu}$
and survives the zero-slope limit (\ref{eq:zero-slope}).
Let us consider the state
\begin{equation}
\zeta_{\mu}(k_{\lambda})b^{\mu}_{-\frac{1}{2}}
  |0;k_{\lambda}\rangle
\otimes
(\overline{\alpha}_{-1+\nu})^{\mn}\,\overline{b}_{-\frac{1}{2}+\nu}
  |\sigma,s\rangle~, \label{eq:state}
\end{equation}
where $\zeta_{\mu}$ is the polarization tensor of this state
and  $|0;k_{\lambda}\rangle$ is the momentum eigenstate
in the $x^{0}$,\ldots,$x^{p}$ directions.
In the above we have divided the state into two pieces;
the former comes from the string coordinates
 $\mathbb{X}^{\mu}$ and the latter from $\mathbb{Z}$ and
 $\overline{\mathbb{Z}}$.
We impose on this state the conditions
\begin{equation}
\alpha^{\prime}m^{2}\equiv -\alpha^{\prime}
                    G^{\mu\lambda}k_{\mu}k_{\lambda}
    =\left(\mn +\frac{1}{2}\right)(1-\nu)~,
 \quad G^{\mu\lambda}k_{\mu}\zeta_{\lambda}=0~.
 \label{eq:scond}
\end{equation}
We find that,
owing to these conditions, the state (\ref{eq:state}) satisfies
the physical state conditions in the $(-1)$-picture.
The on-shell condition in eq.(\ref{eq:scond}) implies that
this state remains light in the zero-slope limit
(\ref{eq:zero-slope}). The state (\ref{eq:state})
is therefore what we wanted.

We denote
the vertex operators whose lower component fields correspond
to the state (\ref{eq:state}) and its conjugate state by
\begin{eqnarray}
&&\mathbb{U}_{\mn}^{(+)}
  (\xi,\theta;k_{\mu};\zeta_{\mu})
 =\mathcal{U}_{\mn}^{(+,-1)}(\xi;k_{\mu};\zeta_{\mu})
 +\theta\,\mathcal{U}_{\mn}^{(+,0)}(\xi;k_{\mu};\zeta_{\mu})~,
\nonumber\\
&&\mathbb{U}_{\mn}^{(-)}
  (\xi,\theta;k_{\mu};\zeta_{\mu})
 =\mathcal{U}_{\mn}^{(-,-1)}(\xi;k_{\mu};\zeta_{\mu})
 +\theta\,\mathcal{U}_{\mn}^{(-,0)}(\xi;k_{\mu};\zeta_{\mu})~,
\end{eqnarray}
respectively. The lower component fields
$\mathcal{U}^{(\pm,-1)}_{\mn}(\xi;k_{\mu};\zeta_{\mu})$
take the forms of
\begin{equation}
  \mathcal{U}^{(\pm,-1)}_{\mn}(\xi;k_{\mu};\zeta_{\mu})
  = \, :\frac{1}{2}\zeta_{\mu}(k_{\lambda})\left(
    \psi^{\mu}+\widetilde{\psi}^{\mu} \right)
    \exp\left(ik_{\rho}X^{\rho}\right)(\xi):
   \, U_{\mn}^{(\pm,-1)}(\xi)~,
\end{equation}
where $U_{\mn}^{(\pm,-1)}(\xi)$ are the contributions from
the fields $\mathbb{Z}$ and $\overline{\mathbb{Z}}$
defined by
\begin{eqnarray}
&&\lim_{\xi\rightarrow 0}U_{\mn}^{(+,-1)}(\xi)\,|0\rangle
 =(\overline{\alpha}_{-1+\nu})^{\mn}
   \,\overline{b}_{-\frac{1}{2}+\nu} \,|\sigma,s\rangle~,\nonumber\\
&&\lim_{\xi\rightarrow\infty}\xi^{2h}\langle 0|\,U_{\mn}^{(-,-1)}(\xi)
 =\langle \sigma,s|\,(\alpha_{1-\nu})^{\mn}\,b_{\frac{1}{2}-\nu}~,
\end{eqnarray}
and $h$ is the conformal weight of $U_{\mn}^{(\pm,-1)}(\xi)$:
$h=\left(\mn+\frac{1}{2}\right)(1-\nu)$.

\subsection{Three point amplitude}

In superstring theory
the gauge field emission vertex operator on the D$(p+2)$-brane
worldvolume is
\begin{equation}
\mathbb{V}_{A}(\xi,\theta;k_{M};\zeta_{M})
=\,:\frac{i}{2}\zeta_{M}(k)\dot{\mathbb{X}}^{M}
 \exp\left[i\sqrt{\frac{\alpha^{\prime}}{2}}
           k_{N}\mathbb{X}^{N} \right](\xi,\theta):~,
\end{equation}
where $\dot{\mathbb{X}}^{M}(\xi,\theta)\equiv\left.(D+\overline{D})
       \mathbb{X}^{M}(\mathbf{z},\overline{\mathbf{z}})\rule{0em}{2ex}
       \right|_{z=\overline{z}=\xi,\theta=\overline{\theta}}$
and $D=\frac{\partial}{\partial\theta}
      +\theta\frac{\partial}{\partial z}$
and $\overline{D}=\frac{\partial}{\partial\overline{\theta}}
     +\overline{\theta}\frac{\partial}{\partial\overline{z}}$ are
the superspace covariant derivatives.
The physical state conditions for this vertex operator is the
same as those in the bosonic string case.

We evaluate the three point disc amplitude
\begin{eqnarray}
&&\mathcal{A}_{3}= c^{\prime}\int
\frac{\prod_{a=1}^{3}d\xi^{(a)} d\theta^{(a)}}{V_{\mathrm{SCKV}}}
\langle 0|\, \mathbb{U}^{(-)}_{\mm}
              (\xi^{(1)},\theta^{(1)};k^{(1)}_{\mu};\zeta_{\mu}^{(1)})
     \, \mathbb{V}_{A}
         (\xi^{(2)},\theta^{(2)};k^{(2)}_{M};\zeta_{M}^{(2)})\nonumber\\
&&\hspace{18em}\times \,\mathbb{U}^{(+)}_{\mn}
              (\xi^{(3)},\theta^{(3)};k^{(3)}_{\mu};\zeta_{\mu}^{(3)})
   \,|0\rangle~,  \label{eq:s3amp}
\end{eqnarray}
where $V_{\mathrm{SCKV}}$ stands for the gauge volume of the
graded $SL(2,\mathbb{R})$ group generated by the superconformal
Killing vectors on the super worldsheet.
To fix the odd elements of the gauge degrees of freedom,
we set $\theta^{(1)}=\theta^{(3)}=0$.
This gauge choice amounts to factoring out the following
volume element from the integration in eq.(\ref{eq:s3amp}):
\begin{equation}
d^{3}F(\xi^{(1)},\xi^{(2)},\xi^{(3)})
    \, d\theta^{(1)}d\theta^{(3)}\, (\xi^{(1)}-\xi^{(3)})~.
\end{equation}
Thus the three point amplitude (\ref{eq:s3amp}) turns out
to be
\begin{eqnarray}
\lefteqn{
  \mathcal{A}_{3}=c^{\prime}(\xi^{(2)}-\xi^{(3)})(\xi^{(3)}-\xi^{(1)})
  \times}
\nonumber\\
&&
\int d\theta^{(2)}\,
\langle 0| \,\mathcal{U}_{\mm}^{(-,-1)}
       (\xi^{(1)};k^{(1)}_{\mu};\zeta^{(1)}_{\mu})
 \;\mathbb{V}_{A}(\xi^{(2)},\theta^{(2)};k^{(2)}_{M};\zeta^{(2)}_{M})
 \;\mathcal{U}_{\mn}^{(+,-1)}(\xi^{(3)};k^{(3)}_{\mu};\zeta^{(3)}_{\mu})
 \,|0\rangle~.~~~~~\label{eq:samp-2}
\end{eqnarray}

As a result of the calculation presented in appendix \ref{sec:scal},
we find that the amplitude $\mathcal{A}_{3}$ becomes
\begin{eqnarray}
&&\mathcal{A}_{3}=c^{\prime}\prod_{\mu=0}^{p}
   \delta(k^{(1)}_{\mu}+k^{(2)}_{\mu}+k^{(3)}_{\mu})
   \prod_{1\leq c<d\leq 3}\exp\left[\frac{i}{2}\theta^{\mu\nu}
  k^{(c)}_{\mu}k^{(d)}_{\lambda}\epsilon(\tau^{(c)}-\tau^{(d)})\right]
   \nonumber\\
 && \hspace{3em}\times 
   \sqrt{\mm!\mn!}
   \left(\frac{2(1-\nu)}{\varepsilon}\right)^{\frac{\mm+\mn}{2}}
   \frac{2}{\varepsilon} \sqrt{\frac{\alpha^{\prime}}{2}}
   \;\mathcal{M}_{\mn\mm}~, \label{eq:samp-3}
\end{eqnarray}
where the factor $\mathcal{M}_{\mn\mm}$ is defined for
the $\mm\geq\mn$ case as
\begin{eqnarray}
&&\hspace{-2em}\mathcal{M}_{\mn\leq\mm}=\sqrt{\frac{\mn!}{\mm!}}
\exp\mathcal{C}(\nu;k^{(2)}_{m})\Bigg[
 -2\left\{ (k^{(2)}\pdot\zeta^{(1)})(\zeta^{(2)}\pdot \zeta^{(3)})
   +(k^{(1)}\pdot\zeta^{(3)})(\zeta^{(1)}\pdot \zeta^{(2)})
 \right\}\nonumber\\
&&\hspace{15em}\times
  \left(i\sqrt{2\Xi}\overline{\kappa}^{(2)}\right)^{\mm-\mn}
 L^{(\mm-\mn)}_{\mn}\left(2\Xi\kappa^{(2)}\overline{\kappa}^{(2)}\right)
 \nonumber\\
 &&+(\zeta^{(3)}\pdot\zeta^{(1)})
   \Bigg\{  \left(-(k^{(3)}-k^{(1)})\pdot \zeta^{(2)}
   +G^{Z\overline{Z}}(\kappa^{(2)}\overline{e}^{(2)}
                    -\overline{\kappa}^{(2)}e^{(2)})\right)
   \nonumber\\
 &&\hspace{15em}
   \times \left(i\sqrt{2\Xi}\overline{\kappa}^{(2)}\right)^{\mm-\mn}
 L^{(\mm-\mn)}_{\mn}\left(2\Xi\kappa^{(2)}\overline{\kappa}^{(2)}\right)
\nonumber\\
&& \hspace{4em} +2G^{Z\overline{Z}}(\kappa^{(2)}\overline{e}^{(2)}
    -\overline{\kappa}^{(2)}e^{(2)})
    \left(i\sqrt{2\Xi}\overline{\kappa}^{(2)}\right)^{\mm-\mn}
L^{(\mm-\mn+1)}_{\mn-1}\left(2\Xi\kappa^{(2)}\overline{\kappa}^{(2)}\right)
\nonumber\\
&&\hspace{4em} -i(\mm-\mn)\sqrt{2\Xi}\frac{1-\nu}{\alpha^{\prime}}
  \overline{e}^{(2)}
  \left(i\sqrt{2\Xi}\overline{\kappa}^{(2)}\right)^{\mm-\mn-1}
  L^{(\mm-\mn)}_{\mn}\left(2\Xi\kappa^{(2)}\overline{\kappa}^{(2)}\right)
  \Bigg\}\Bigg]~,
\end{eqnarray}
and for the $\mm<\mn$ case as
\begin{eqnarray}
&&\hspace{-2em}\mathcal{M}_{\mn\geq\mm}=\sqrt{\frac{\mm!}{\mn!}}
\exp\mathcal{C}(\nu;k^{(2)}_{m})\Bigg[
 -2\left\{ (k^{(2)}\pdot\zeta^{(1)})(\zeta^{(2)}\pdot \zeta^{(3)})
   +(k^{(1)}\pdot\zeta^{(3)})(\zeta^{(1)}\pdot \zeta^{(2)})
 \right\}\nonumber\\
&&\hspace{15em}\times
  \left(i\sqrt{2\Xi}\kappa^{(2)}\right)^{\mn-\mm}
 L^{(\mn-\mm)}_{\mm}\left(2\Xi\kappa^{(2)}\overline{\kappa}^{(2)}\right)
 \nonumber\\
 &&+(\zeta^{(3)}\pdot\zeta^{(1)})
   \Bigg\{  \left(-(k^{(3)}-k^{(1)})\pdot \zeta^{(2)}
   +G^{Z\overline{Z}}(\kappa^{(2)}\overline{e}^{(2)}
                    -\overline{\kappa}^{(2)}e^{(2)})\right)
   \nonumber\\
 &&\hspace{15em}
   \times \left(i\sqrt{2\Xi}\kappa^{(2)}\right)^{\mn-\mm}
 L^{(\mn-\mm)}_{\mn}\left(2\Xi\kappa^{(2)}\overline{\kappa}^{(2)}\right)
\nonumber\\
&& \hspace{4em} +2G^{Z\overline{Z}}(\kappa^{(2)}\overline{e}^{(2)}
    -\overline{\kappa}^{(2)}e^{(2)})
    \left(i\sqrt{2\Xi}\kappa^{(2)}\right)^{\mn-\mm}
L^{(\mn-\mm+1)}_{\mm-1}\left(2\Xi\kappa^{(2)}\overline{\kappa}^{(2)}\right)
\nonumber\\
&&\hspace{4em} +i(\mn-\mm)\sqrt{2\Xi}\frac{1-\nu}{\alpha^{\prime}}
  e^{(2)}
  \left(i\sqrt{2\Xi}\kappa^{(2)}\right)^{\mn-\mm-1}
  L^{(\mn-\mm)}_{\mm}\left(2\Xi\kappa^{(2)}\overline{\kappa}^{(2)}\right)
  \Bigg\}\Bigg]~.
\end{eqnarray}
Comparing the above equations with
eqs.(\ref{eq:kln<m}) and (\ref{eq:kln>m}),
we find that the factor $\mathcal{M}_{\mn\mm}$ consists of
terms which quite resemble  $\mathcal{K}_{\mn\mm}$.
We can therefore rewrite the above equations into the form
involving the factors $\mathcal{I}_{\mn\mm}$
as we carried out in eqs.(\ref{eq:kn<m}) and (\ref{eq:kn>m}).
Here we focus on $\mathcal{M}_{\mm\mm}$ in particular:
\begin{eqnarray}
\lefteqn{\mathcal{M}_{\mm\mm}=-2\left\{
  (k^{(2)}\pdot\zeta^{(1)})(\zeta^{(2)}\pdot\zeta^{(3)})
  +(k^{(1)}\pdot\zeta^{(3)})(\zeta^{(1)}\pdot\zeta^{(2)})
  \right\}\mathcal{I}_{\mm\mm}}\nonumber\\
&&\hspace{2em}+(\zeta^{(3)}\pdot\zeta^{(1)})
  \Big\{ \left(-(k^{(3)}-k^{(1)})\pdot \zeta^{(2)}
   +G^{Z\overline{Z}}(\kappa^{(2)}\overline{e}^{(2)}
                    -\overline{\kappa}^{(2)}e^{(2)})\right)
  \mathcal{I}_{\mm\mm}\nonumber\\
&&\hspace{8em}+2G^{Z\overline{Z}}(\kappa^{(2)}\overline{e}^{(2)}
  -\overline{\kappa}^{(2)}e^{(2)}) \sum_{\mathrm{k}=0}^{\mm-1}
  \mathcal{I}_{\mathrm{kk}}\Big\}~.
\end{eqnarray}
{}From this equation, we find that in the case of $\mm=\mn$
the amplitude involves the multiplicative factors
$\mathcal{I}_{\mm\mm}$ and
$\sum_{\mathrm{k}=0}^{\mm-1}\mathcal{I}_{\mathrm{kk}}$
which become projection operators $|\mm\rangle\langle \mm|$
and $\sum_{\mathrm{k}=0}^{\mm-1}|\mathrm{k}\rangle\langle\mathrm{k}|$
respectively in the zero-slope limit.
These are what we desired.

\section{Conclusion and Discussions}

In this paper we have calculated three point scattering amplitudes
for processes probing the D$p$-brane with the $p^{\prime}$-$p^{\prime}$
open string
in the $p$-$p^{\prime}$ system with $B$ field.
We carried out the calculation
in bosonic string theory and in superstring theory.
We have focused on  processes in which the almost-zero-modes
of the $p$-$p^{\prime}$ open string are involved.
Following ref.~\cite{CIMM2},
we have read off momentum dependent multiplicative factors from
three point disc amplitudes for such processes.
We have showed that in the case of $\mm=\mn$ these multiplicative
factors take such forms that become Weyl transforms
of projection operators in the zero-slope limit.

We think that this result provides further evidence
for the observation that the tower of the light states
surviving the zero slope limit \cite{SW}\cite{CIMM1}
play a role to realize
the D-branes as noncommutative solitons
on the higher dimensional
noncommutative D-brane worldvolume.

The fact that many light states survive
the zero-slope limit
is a particular result of a nonvanishing $B$ field background.
Thus it is expected that the tower of light states
cause the behavior of D-branes within a  noncommutative
D-brane to look quite different from that within
a commutative D-brane.
The roles played by the tower of the light states in this system
seem to need more investigation and clarification.
As well as the perturbative string analysis,
the field theoretical analysis
is expected to be important and useful in some aspect
of this problem.

\section*{Acknowledgements}
The author would like to thank to H.~Itoyama,
H.~Kunitomo, G.~Mandal,
T.~Matsuo and T.~Suyama for discussions
and comments.
This work is supported in part by JSPS Research Fellowships
for Young Scientists.


\appendix

\section{Weyl Correspondence}\label{sec:Weylcorr}

There is a one-to-one correspondence,
referred to as the Weyl correspondence, between a function
on the commutative $\mathbb{R}^{2}$ space and an operator
on noncommutative $\mathbb{R}^{2}$ space
(see e.g.\ \cite{GMS}\cite{komaba}).
In order to fix the notation, we would like to give some formulae
of Weyl ordering prescription.

Let us consider the noncommutative $\mathbb{R}^{2}$ space
characterized by\footnote{The coordinates $x^{1}$ and $x^{2}$ in this
appendix correspond to $x^{p+1}$ and $x^{p+2}$
in the main part of this paper respectively.
The minus sign on the r.h.s.\ in eq.(\ref{eq:NCR2}) is necessary
because we consider the $\sigma=\pi$ end of the string \cite{CH}
(i.e.\ the negative real axis on the $z$-plane of the worldsheet).
This is just a convention.}
\begin{equation}
[\hat{x}^{1},\hat{x}^{2}]=-i\theta^{12}~,
\quad \theta^{12}\in\mathbb{R}~.  \label{eq:NCR2}
\end{equation}
This allows us to regard the noncommutative $\mathbb{R}^{2}$ space
as the single particle Hilbert space $\mathcal{H}$ of quantum mechanics.
Given a function $f(x^{1},x^{2})$ on the commutative $\mathbb{R}^{2}$ space,
we uniquely determine a corresponding operator
$\widehat{\mathcal{O}}_{f}(\hat{x}^{1},\hat{x}^{2})$
on the noncommutative $\mathbb{R}^{2}$ space
by the Weyl ordering prescription:
\begin{eqnarray}
&&\widehat{\mathcal{O}}_{f}(\hat{x}^{1},\hat{x}^{2})
\equiv \int \frac{dk_{1}dk_{2}}{(2\pi)^{2}}
\widehat{V}(k_{1},k_{2}) \widetilde{f}(k_{1},k_{2})~,
\nonumber\\
 && \qquad \widehat{V}(k_{1},k_{2}) \equiv
  \exp\left[-i\left(k_{1}\hat{x}^{1}+k_{2}\hat{x}^{2}\right)\right]~,
  \label{eq:ovf}
\end{eqnarray}
where $\widetilde{f}(k_{1},k_{2})$ is the Fourier transform
of the function $f(x^{1},x^{2})$:
\begin{equation}
\widetilde{f}(k_{1},k_{2})\equiv \int dx^{1}dx^{2}
  e^{ i\left(k_{1}x^{1}+k_{2}x^{2}\right)}  f(x^{1},x^{2})~.
\end{equation}
Or equivalently,
\begin{eqnarray}
&&\widehat{\mathcal{O}}_{f}(\hat{x}^{1},\hat{x}^{2})
 =\int dx^{1}dx^{2}\widehat{\triangle}(x^{1},x^{2})f(x^{1},x^{2})~,
 \nonumber\\
&& \qquad \widehat{\triangle}(x^{1},x^{2})
  \equiv \int \frac{dk_{1}dk_{2}}{(2\pi)^{2}}
  \exp \left[-ik_{1}(\hat{x}^{1}-x^{1})-ik_{2}(\hat{x}^{2}-x^{2})\right]~.
  \label{eq:odf}
\end{eqnarray}
The inverse transformation of eq.(\ref{eq:odf}) is
\begin{eqnarray}
f(x^{1},x^{2})&=&|\theta^{12}|\int dk_{2}\,e^{-ik_{2}x^{2}}
 \left\langle x^{1}-\frac{\theta^{12}}{2}k_{2}\right|
  \widehat{\mathcal{O}}_{f}(\hat{x}^{1},\hat{x}^{2})
 \left| x^{1}+\frac{\theta^{12}}{2}k_{2}\right\rangle\nonumber\\
 &=&|\theta^{12}|\int dk_{1}\, e^{-ik_{1}x^{1}}
  \left\langle x^{2}+\frac{\theta^{12}}{2}k_{1}\right|
   \widehat{\mathcal{O}}_{f}(\hat{x}^{1},\hat{x}^{2})
  \left| x^{2}-\frac{\theta^{12}}{2}k_{1}\right\rangle~,
\end{eqnarray}
where $|x^{1}\rangle$ and $|x^{2}\rangle$ are the eigenstates
of the operators $\hat{x}^{1}$ and $\hat{x}^{2}$ respectively:
$\hat{x}^{1}|x^{1}\rangle =x^{1}|x^{1}\rangle$,
$\hat{x}^{2}|x^{2}\rangle =x^{2}|x^{2}\rangle$, normalized as
$\langle x^{\prime 1}|x^{1}\rangle=\delta(x^{\prime 1}-x^{1})$
and $\langle x^{\prime 2}|x^{2}\rangle=\delta(x^{\prime 2}-x^{2})$.
This can be proved by using the relations
\begin{eqnarray}
\widehat{\triangle}(\hat{x}^{1},\hat{x}^{2})
 &=&\int\frac{dk_{2}}{2\pi}\, e^{ik_{2}x^{2}}
  \left|x^{1}-\frac{\theta^{12}}{2}k_{2}\right\rangle
       \left\langle x^{1}+\frac{\theta^{12}}{2}k_{2}\right|
      \nonumber\\
  &=& \int\frac{dk_{1}}{2\pi}\, e^{ik_{1}x^{1}}
    \left|x^{2}+\frac{\theta^{12}}{2}k_{1}\right\rangle
       \left\langle x^{2}-\frac{\theta^{12}}{2}k_{1}\right|~.
\end{eqnarray}
It is well-known that this correspondence relation
maps the product of the operators
into the $\ast$-product of the functions:
$\widehat{\mathcal{O}}_{f}\widehat{\mathcal{O}}_{g}
=\widehat{\mathcal{O}}_{f\ast g}$, where
\begin{equation}
f\ast g (x^{1},x^{2})
 \equiv f(x^{1},x^{2}) e^{-i\theta^{mn}\stackrel{\leftarrow}{\partial}_{m}
                                 \stackrel{\rightarrow}{\partial}_{n}}
  g(x^{1},x^{2})~.
\end{equation}

The formula
\begin{equation}
\mathrm{Tr}_{\mathcal{H}} \left( \widehat{V}^{\dagger}(u_{1},u_{2})
                   \widehat{V}(k_{1},k_{2})\right)
       =\frac{2\pi}{|\theta^{12}|}\delta(u_{1}-k_{1})\delta(u_{2}-k_{2})
\end{equation}
and eq.(\ref{eq:ovf}) yield
\begin{equation}
\widetilde{f}(k_{1},k_{2})=2\pi |\theta^{12}|\ %
 \mathrm{Tr}_{\mathcal{H}}\left( \widehat{V}^{\dagger}(k_{1},k_{2})
  \widehat{\mathcal{O}}_{f}(\hat{x}^{1},\hat{x}^{2})\right)~.
  \label{eq:trvv}
\end{equation}

Let us define the creation operator $\hat{a}^{\dagger}$
and the annihilation operator $\hat{a}$.
We consider the case
\footnote{For the case where $\theta^{12}>0$,
    we should exchange the space indices $1$ and $2$ in
    the subsequent equations. For example,
    $\hat{a}$ and $\hat{a}^{\dagger}$ would be defined as
    $\hat{a}\equiv
      \frac{\hat{x}^{2}+i\hat{x}^{1}}{\sqrt{2|\theta^{12}|}}$,
    $\hat{a}^{\dagger}\equiv
       \frac{\hat{x}^{2}-i\hat{x}^{1}}{\sqrt{2|\theta^{12}|}}$.}
in which $\theta^{12}<0$ and $[\hat{x}^{1},\hat{x}^{2}]=i|\theta^{12}|$.
This is the situation
analyzed in the main part of this paper
(see eq.(\ref{eq:theta})).
In this case we define $\hat{a}$ and $\hat{a}^{\dagger}$ as
\begin{equation}
\hat{a}\equiv \frac{\hat{x}^{1}+i\hat{x}^{2}}{\sqrt{2|\theta^{12}|}}~,
\qquad \hat{a}^{\dagger}\equiv
       \frac{\hat{x}^{1}-i\hat{x}^{2}}{\sqrt{2|\theta^{12}|}}~.
    \label{eq:aad}
\end{equation}

Let $f_{\mn\mm}(x^{1},x^{2})$ be the Weyl transform of the operator
$| \mn \rangle\langle \mm |$ $(\mm,\mn=0,1,2,\ldots)$,
where $|\mm\rangle$ are the harmonic oscillator eigenstates:
$ |\mm\rangle
   \equiv \frac{(\hat{a}^{\dagger})^{\mm}}{\sqrt{\mm!}}|0\rangle$.
{}From eq.(\ref{eq:trvv}) one can find that the Fourier transform
$\widetilde{f}_{\mn\mm}(k_{1},k_{2})$ of the function
$f_{\mn\mm}(x^{1},x^{2})$ takes the form of
\begin{equation}
\widetilde{f}_{\mn\mm}(k_{1},k_{2})=2\pi|\theta^{12}|
\ \mathrm{Tr}_{\mathcal{H}}\left( \widehat{V}^{\dagger}(k_{1},k_{2})
    |\mn\rangle\langle \mm| \right)
  =2\pi |\theta^{12}| \, \langle \mm| \widehat{V}^{\dagger}(k_{1},k_{2})
        |\mn \rangle~.
\end{equation}
We can rewrite the operator $\widehat{V}^{\dagger}(k_{1},k_{2})$ as
\begin{equation}
\widehat{V}^{\dagger}(k_{1},k_{2})=
e^{i\sqrt{2|\theta^{12}|}\;
   \left(\kappa\hat{a}+\overline{\kappa}\hat{a}^{\dagger}\right)}
= e^{-|\theta^{12}|\kappa\overline{\kappa}}
\,e^{i\sqrt{2|\theta^{12}|}\;
           \overline{\kappa}\hat{a}^{\dagger}}
\,e^{i\sqrt{2|\theta^{12}|}\;
           \kappa \hat{a}}~,
\end{equation}
where $\kappa=\frac{1}{2}(k_{1}-ik_{2})$ and
$\overline{\kappa}=\frac{1}{2}(k_{1}+ik_{2})$.
This yields
\begin{eqnarray}
\lefteqn{ \widetilde{f}_{\mn\mm}(k_{1},k_{2})} \nonumber\\
&&=\left\{
 \begin{array}{ll}
  \displaystyle
    2\pi |\theta^{12}| \sqrt{\frac{\mn!}{\mm!}}
    \left(i\sqrt{2|\theta^{12}|}\;\overline{\kappa} \right)^{\mm-\mn}
      e^{-|\theta^{12}|\kappa\overline{\kappa}}
     L^{(\mm-\mn)}_{\mn} 
       \Big(2 |\theta^{12}|\kappa\overline{\kappa} \Big)
  & \ (\mm\geq \mn)\\[2ex]
  \displaystyle
    2\pi |\theta^{12}| \sqrt{\frac{\mm!}{\mn!}}
     \left(i\sqrt{2|\theta^{12}|}\;\kappa\right)^{\mn-\mm}
     e^{-|\theta^{12}|\kappa\overline{\kappa}}
     L^{(\mn-\mm)}_{\mm}\Big(2 |\theta^{12}|\kappa\overline{\kappa}\Big)
    & \ (\mm<\mn)
 \end{array} \right. \!\!,  \label{eq:weylnm}
\end{eqnarray}
where $L^{(\alpha)}_{\mn}(x)$ is
the Laguerre polynomial (\ref{eq:Laguerre}).

\section{Calculation of $\mathcal{A}_{3}^{\mathrm{tachyon}}$}
\label{sec:a3tach}

In this appendix we give some details in the
calculation of eq.(\ref{eq:a3-f}).

The location $\xi^{(2)}$ of the tachyon vertex operator
$\mathcal{V}_{\phi}$ is restricted to the
interval $\xi^{(2)}\in [\xi^{(1)},\xi^{(3)}]$ as explained.
Using the $SL(2,\mathbb{R})$ gauge degrees of freedom
on the worldsheet,
we choose $\xi^{(1)}=-\infty$ and $\xi^{(3)}=0$,
so that the negative real axis becomes the worldsheet boundary
attaching on the D$(p+2)$-brane.
The correlation function in the integrand on the r.h.s.\ in
eq.(\ref{eq:a3}) is factorized into two pieces;
the one is the contribution from $X^{0},\ldots, X^{p}$
and the other is form $Z$ and $\overline{Z}$:
\begin{eqnarray}
\lefteqn{\langle 0| \mathcal{V}^{(-)}_{\mm}(\xi^{(1)};k_{\mu}^{(1)})
           \,\mathcal{V}_{\phi}(\xi^{(2)};k_{M}^{(2)})
           \,\mathcal{V}^{(+)}_{\mn}(\xi^{(3)};k_{\mu}^{(3)}) |0\rangle}
      \nonumber\\
   &&     =\langle 0|
    :\exp\left[ ik^{(1)}_{\mu}X^{\mu}\right](\xi^{(1)}):
    :\exp\left[ ik^{(2)}_{\mu}X^{\mu}\right] (\xi^{(2)}):
    :\exp\left[ik^{(3)}_{\mu}X^{\mu}\right] (\xi^{(3)}):
    |0\rangle\nonumber\\
&& \hspace{2em}
\times \langle 0| V^{(-)}_{\mm}(\xi^{(1)})
          \,V_{\phi}(\xi^{(2)};k^{(2)}_{m})
          \,V^{(+)}_{\mn}(\xi^{(3)}) |0\rangle~. \label{eq:3pt-ta}
\end{eqnarray}
The propagator (\ref{eq:sw2pt-b}) leads us to find that the former
piece becomes \cite{Jabbari}\cite{Schomerus}\cite{SW}
\begin{eqnarray}
\label{eq:mphin}
&&\langle 0 |
  :\exp\left[ik^{(1)}_{\mu}X^{\mu}\right](\xi^{(1)}):\,
    :\exp\left[ik^{(2)}_{\mu}X^{\mu}\right] (\xi^{(2)}):
    :\exp\left[ik^{(3)}_{\mu}X^{\mu}\right](\xi^{(3)}):
    |0\rangle \\ 
&& =\prod_{\mu=0}^{p}\delta(k^{(1)}_{\mu}+k^{(2)}_{\mu}+k^{(3)}_{\mu})
   \prod_{1\leq c<d \leq 3}
 (x^{(c)}-x^{(d)})^{2\alpha^{\prime}G^{\mu\nu}k^{(c)}_{\mu}k^{(d)}_{\nu}}
 \exp\left[\frac{i}{2}\theta^{\mu\nu}k_{\mu}^{(c)}k_{\nu}^{(d)}
    \epsilon(x^{(c)}-x^{(d)})\right]~.\nonumber
\end{eqnarray}
Here we have introduced positive real variables
$x^{(c)}\equiv -\xi^{(c)}=e^{\tau^{(c)}}$
$(c=1,2,3)$.
Since we have chosen $x^{(1)}=\infty$ and $x^{(3)}=0$,
the latter piece on the r.h.s.\ of eq.(\ref{eq:3pt-ta})
becomes
\begin{eqnarray}
\lefteqn{
  \langle 0| V^{(-)}_{\mm}(-x^{(1)})\,V_{\phi}(-x^{(2)};k^{(2)}_{m})
               \,V^{(+)}_{\mn}(-x^{(3)}) |0\rangle}\nonumber\\
&&  =\left(x^{(1)}\right)^{-(2\mm+\nu)(1-\nu)}
   \left(x^{(2)}\right)^{-2\alpha^{\prime}G^{Z\overline{Z}}
           \kappa^{(2)}\overline{\kappa}^{(2)}}
    \exp{\mathcal{C}\left(\nu;k^{(2)}_{m}\right)}\nonumber\\
&&\hspace{1.5em}\times
     \langle\sigma|\left(\alpha_{1-\nu}\right)^{\mm}
  \maru~\exp\left[i\left(\kappa^{(2)}Z+\overline{\kappa}^{(2)}\overline{Z}
            \right)\right](-x^{(2)})~\maru
      \left(\overline{\alpha}_{-1+\nu}\right)^{\mn}|\sigma\rangle~,
    \label{eq:3tach}
\end{eqnarray}
with $x^{(1)}=\infty$ and $x^{(3)}=0$,
where $\mathcal{C}(\nu;k_{m})$ is defined in eq.(\ref{eq:cfactor}).
Here we have used the defining relations (\ref{eq:vnm})
and applied the renormal ordering formula~(\ref{eq:renormal})
to the operator $V_{\phi}(-x^{(2)};k^{(2)}_{m})$:
\begin{eqnarray}
V_{\phi}(-x^{(2)};k^{(2)}_{m})
   =\left(x^{(2)}\right)^{-2\alpha^{\prime}G^{Z\overline{Z}}
                          \kappa^{(2)}\overline{\kappa}^{(2)}}
    \exp\mathcal{C}(\nu;k^{(2)}_{m})
    \,\,\maru\exp\left[i\left(\kappa^{(2)} Z
                   +\overline{\kappa}^{(2)}\overline{Z}\right)\right]
                   (-x^{(2)})\maru~.
\end{eqnarray}
Now that the operator
$\maru~\exp\left[i\left(\kappa^{(2)} Z
    +\overline{\kappa}^{(2)}\overline{Z}\right)\right](-x^{(2)})~\maru$
is evaluated on the oscillator vacuum $|\sigma\rangle$,
the fields $Z$ and $\overline{Z}$ should be expanded
as is described in eq.(\ref{eq:Zexp})
and the normal ordering defined on $|\sigma\rangle$ in \cite{CIMM2}
be taken.
Thus, by using the commutation relation (\ref{eq:com}), we obtain
\begin{eqnarray}
 [\,\overline{\alpha}_{-1+\nu}~,~\maru~e^{i\left(\kappa^{(2)} Z
    +\overline{\kappa}^{(2)}\overline{Z}\right)}(-x^{(2)})~\maru
    \,]
  &=&f\left(\nu;x^{(2)};\kappa^{(2)}\right)
     \,\maru~e^{i\left(\kappa^{(2)} Z
       +\overline{\kappa}^{(2)}\overline{Z}\right)}(-x^{(2)})~\maru~,
 \nonumber\\
{}[\,\alpha_{1-\nu}~,~\maru~e^{i\left(\kappa^{(2)} Z
    +\overline{\kappa}^{(2)}\overline{Z}\right)}(-x^{(2)})~\maru
    \,]
   &=&g\left(\nu;x^{(2)};\overline{\kappa}^{(2)}\right)
     \,\maru~e^{i\left(\kappa^{(2)} Z
    +\overline{\kappa}^{(2)}\overline{Z}\right)}(-x^{(2)})~\maru~,~~~
 \label{eq:alV}
\end{eqnarray}
where
\begin{eqnarray}
f\left(\nu;x^{(2)};\kappa^{(2)}\right)
&\equiv& -\left(x^{(2)}\right)^{-1+\nu}
       i\kappa^{(2)}\sqrt{\frac{\alpha^{\prime}}{2}}
        \frac{4}{\varepsilon}
        \sin(\pi \nu)~,\nonumber\\
g\left(\nu;x^{(2)};\overline{\kappa}^{(2)}\right)
&\equiv&\left(x^{(2)}\right)^{1-\nu}i\overline{\kappa}^{(2)}
        \sqrt{\frac{\alpha^{\prime}}{2}}\frac{4}{\varepsilon}
        \sin (\pi\nu)~.
\end{eqnarray}
The commutation relations (\ref{eq:alV})
and $[\alpha_{1-\nu},\overline{\alpha}_{-1+\nu}]
      =\frac{2}{\varepsilon}(1-\nu)(\equiv q)$ yield
\begin{eqnarray}
\lefteqn{ \left(\alpha_{1-\nu}\right)^{\mm}
 \maru~\exp\left[i\left(\kappa^{(2)}Z+\overline{\kappa}^{(2)}\overline{Z}
            \right)\right](-x^{(2)})~\maru
 \left(\overline{\alpha}_{-1+\nu}\right)^{\mn}}\nonumber\\
&&=\left\{
 \begin{array}{ll}
 \displaystyle
   \sum_{l=0}^{\mn}
    \frac{\mm!\; \mn!}{l!(\mn-l)!\,(l+\mm-\mn)!}q^{\mn-l}
    \left(\overline{\alpha}_{-1+\nu}-f\right)^{l}
    \left(g+\alpha_{1-\nu}\right)^{\mm-\mn+l}
    & \mbox{($\mm\geq \mn$)}\\
  \displaystyle
   \sum_{l=0}^{\mm} \frac{\mm!\; \mn!}{l!(\mm-l)!\,(l+\mn-\mm)!}
    q^{\mm-l}\left(\overline{\alpha}_{-1+\nu}-f\right)^{l+(\mn-\mm)}
    \left(g+\alpha_{1-\nu}\right)^{l}
    & \mbox{($\mm<\mn$)}
 \end{array}
\right.\!\!\!.~~~~
\end{eqnarray}
{}From this we find that
\begin{eqnarray}
 &&
  \langle \sigma | \left(\alpha_{1-\nu}\right)^{\mm}
 \maru~\exp\left[i\left(\kappa^{(2)}Z+\overline{\kappa}^{(2)}\overline{Z}
            \right)\right](-x^{(2)})~\maru
 \left(\overline{\alpha}_{-1+\nu}\right)^{\mn}
| \sigma\rangle
\nonumber\\
&&\hspace{5em}
=\left\{
 \begin{array}{ll}
  \displaystyle \mn!\,q^{\mn}g^{\mm-\mn}\,L_{\mn}^{(\mm-\mn)}
    \left(\frac{fg}{q}\right)
  &\quad \mbox{($\mm\geq \mn$)}\\[3ex]
  \displaystyle \mm! \, q^{\mm} (-f)^{\mn-\mm}\,
     L_{\mm}^{(\mn-\mm)}\left(\frac{fg}{q}\right)
  &\quad\mbox{($\mm<\mn$)}
 \end{array}
\right.~,     \label{eq:mVn}
\end{eqnarray}
where
$L^{(\alpha)}_{\mn}(x)$ is
the Laguerre polynomial (\ref{eq:Laguerre}).
Substituting eqs.(\ref{eq:mphin}), (\ref{eq:3tach}) and (\ref{eq:mVn})
into eq.(\ref{eq:3pt-ta})
and using the on-shell conditions,
we obtain the three point amplitude (\ref{eq:a3-f}).

\section{Calculation of $\mathcal{A}_{3}^{\mathrm{gauge}}$}
\label{sec:a3gauge}

In this appendix we provide the steps to obtain
eq.(\ref{eq:a3vec-f}).

Introducing a parameter $a$, we rewrite the vertex operator
into the following form \cite{IM}:
\begin{eqnarray}
\lefteqn{
   \mathcal{V}_{A}(\xi;k_{M};\zeta_{M})=\frac{\partial}{\partial a}
 \left.
 : \exp \left[ i\left(k_{M}X^{M}+a\zeta_{M}\dot{X}^{M}\right)
        \right](\xi):
 \;\right|_{a=0}} \nonumber\\
   &&= \left. \frac{\partial}{\partial a} \left\{
     : \exp \left[ i\left(k_{\mu}X^{\mu}+a\zeta_{\mu}\dot{X}^{\mu} \right)
            \right]:
     : \exp \left[ i \left(\kappa Z+\overline{\kappa}\overline{Z}
                    +ae\dot{Z}                   
                    +a\overline{e}\dot{\overline{Z}}\right) \right]
       :\right\} \,  \right|_{a=0}~,~~
\end{eqnarray}
where $e$ and $\overline{e}$ are defined in eq.(\ref{eq:eebar}).
As was done in the appendix \ref{sec:a3tach},
we divide the correlation function in the integrand
on the r.h.s.\ in eq.(\ref{eq:a3vec}) into two pieces;
the contribution from $X^{\mu}$
$(\mu=0,\ldots,p)$ and that from $Z$ and $\overline{Z}$:
\begin{eqnarray}
&&\hspace{-2em}
   \langle 0| \mathcal{V}_{\mm}^{(-)}(\xi^{(1)};k^{(1)}_{\mu})\,
   \mathcal{V}_{A}(\xi^{(2)};k^{(2)}_{M};\zeta_{M})\,
   \mathcal{V}_{\mn}^{(+)}(\xi^{(3)};k^{(3)}_{\mu})
   |0\rangle\nonumber\\
&& \hspace{-1.5em}=
\frac{d}{da}\left[
 \langle 0| :\exp \left(k^{(1)}_{\mu}X^{\mu}\right)(\xi^{(1)}):
  \; :\exp\left[ i \left(k^{(2)}_{\mu}X^{\mu}+a\zeta_{\mu}\dot{X}^{\mu}
                   \right) \right](\xi^{(2)}):
  \; :\exp\left(ik^{(3)}_{\mu}X^{\mu}\right)(\xi^{(3)}): 
 |0\rangle \right.\nonumber\\
&&\left. \ \times \langle 0 | \, V^{(-)}_{\mm}(\xi^{(1)})\;
     :\exp\left[ i\left(\kappa^{(2)}Z+\overline{\kappa}^{(2)}\overline{Z}
             +ae\dot{Z}+a\overline{e}\dot{\overline{Z}} \right) \right]
             (\xi^{(2)}):
   \;  V_{\mn}^{(+)}(\xi^{(3)}) \, |0\rangle
   \right]_{a=0}~.~\label{eq:2gauge}
\end{eqnarray}
The former piece in eq.(\ref{eq:2gauge}) becomes
\cite{SW}\cite{Jabbari}\cite{Schomerus}\cite{IM}
\begin{eqnarray}
\lefteqn{\langle 0|
      :\exp \left(k^{(1)}_{\mu}X^{\mu}\right)(-x^{(1)}):
  \; :\exp\left[ i \left(k^{(2)}_{\mu}X^{\mu}+a\zeta_{\mu}\dot{X}^{\mu}
                   \right) \right](-x^{(2)}):
  \; :\exp\left(ik^{(3)}_{\mu}X^{\mu}\right)(-x^{(3)}): 
  |0 \rangle}\nonumber\\
&&
 = \prod_{\mu=0}^{p}\delta(k^{(1)}_{\mu}+k^{(2)}_{\mu}+k^{(3)}_{\mu})
 \prod_{1\leq c<d \leq 3}
 (x^{(c)}-x^{(d)})^{2\alpha^{\prime}G^{\mu\nu}k^{(c)}_{\mu}k^{(d)}_{\nu}}
 \exp\left[\frac{i}{2}\theta^{\mu\nu}k^{(c)}_{\mu}k^{(d)}_{\nu}
           \epsilon(\tau^{(c)}-\tau^{(d)}) \right] \nonumber\\
&&\hspace{1.5em}\times
   \exp \left[a  \left(
     \frac{2\alpha^{\prime}G^{\mu\nu}k^{(1)}_{\mu}\zeta_{\nu}}
          {x^{(1)}-x^{(2)}}
     - \frac{2\alpha^{\prime}G^{\mu\nu}k^{(3)}_{\mu}\zeta_{\nu}}
          {x^{(2)}-x^{(3)}}\right) \right]~.
  \label{eq:tvt1}
\end{eqnarray}
Sending $x^{(1)}\rightarrow \infty$ and $x^{(3)}\rightarrow 0$,
we find that the latter piece in eq.(\ref{eq:2gauge})
takes the form of
\begin{eqnarray}
\lefteqn{\langle 0 | \, V^{(-)}_{\mm}(-x^{(1)})\;
     :\exp\left[ i\left(\kappa^{(2)}Z+\overline{\kappa}^{(2)}\overline{Z}
             +ae\dot{Z}+a\overline{e}\dot{\overline{Z}} \right) \right]
             (-x^{(2)}):
   \;  V_{\mn}^{(+)}(-x^{(3)}) \, |0\rangle} \nonumber\\  
&&=(x^{(1)})^{-(2\mm+\nu)(1-\nu)}
(x^{(2)})^{-2\alpha^{\prime}G^{Z\overline{Z}}
             \kappa^{(2)}\overline{\kappa}^{(2)}}\nonumber\\
  && \times \exp\left[\mathcal{C}(\nu;k^{(2)}_{m})
              +2\alpha^{\prime}G^{Z\overline{Z}}\left\{
        a\left( \frac{\kappa^{(2)}\overline{e}\nu
                +\overline{\kappa}^{(2)}e(1-\nu)}{x^{(2)}} \right)
        +a^{2}e\overline{e}\frac{\nu(1-\nu)}{2(x^{(2)})^{2}}
       \right\} \right]\nonumber\\
   && \times \langle \sigma |
     (\alpha_{1-\nu})^{\mm}
     \; \maru \exp\left[ i \left(\kappa^{(2)}Z
       +\overline{\kappa}^{(2)}\overline{Z}
       +ae\dot{Z}+a \overline{e}\dot{\overline{Z}} \right)\right]
       (-x^{(2)})\maru
      \; (\overline{\alpha}_{-1+\nu})^{\mn} |\sigma\rangle~,
   \label{eq:tvt2}
\end{eqnarray}
with $x^{(1)}=\infty$ and $x^{(3)}=0$.
Here we have applied the renormal ordering formula (\ref{eq:renormal})
to the gauge field vertex operator.
In a similar way to the calculation of eq.(\ref{eq:mVn}),
we obtain
\begin{eqnarray}
&& \langle \sigma | (\alpha_{1-\nu})^{\mm}
  \;\maru \exp\left[i\left(
    \kappa^{(2)}Z+\overline{\kappa}^{(2)}\overline{Z}
    +a(e\dot{Z}+\overline{e}\dot{\overline{Z}})\right)\right]
    (-x^{(2)})\maru
  \; (\overline{\alpha}_{-1+\nu})^{\mn}|\sigma\rangle
  \nonumber\\
 &&\hspace{3em}=\left\{
  \begin{array}{ll}
   \displaystyle
      \mn!\,q^{\mn}\widetilde{g}^{\mm-\mn}L^{(\mm-\mn)}_{\mn}\left(
         \frac{\widetilde{f}\;\widetilde{g}}{q} \right)
         & \mbox{ ($\mm\geq \mn$)}\\[3ex]
   \displaystyle
      \mm!\,q^{\mm}(-\widetilde{f})^{\mn-\mm}L^{(\mn-\mm)}_{\mm}
      \left(\frac{\widetilde{f}\;\widetilde{g}}{q} \right)
      & \mbox{ ($\mm<\mn$)}
  \end{array}  \right.  , \label{eq:aVavec}
\end{eqnarray}
where
\begin{eqnarray}
  \widetilde{f}(\nu;x^{(2)};\kappa^{(2)};e)
  &\equiv& -(x^{(2)})^{-1+\nu}
    \,i\left\{\kappa^{(2)}+a\frac{(1-\nu)e}{x^{(2)}}\right\}
        \sqrt{\frac{\alpha^{\prime}}{2}}\frac{4}{\varepsilon}
      \sin (\pi \nu)~,  \nonumber\\
   \widetilde{g}(\nu;x^{(2)};\overline{\kappa}^{(2)};\overline{e})
   &\equiv& (x^{(2)})^{1-\nu}\,i
     \left\{\overline{\kappa}^{(2)}
       -a\frac{(1-\nu)\overline{e}}{x^{(2)}} \right\}
     \sqrt{\frac{\alpha^{\prime}}{2}} \frac{4}{\varepsilon}
     \sin (\pi\nu)~.
\end{eqnarray}
Plugging eqs.(\ref{eq:tvt1}), (\ref{eq:tvt2}) and (\ref{eq:aVavec})
into eq.(\ref{eq:a3vec})
and using the physical state conditions and the relation
for the Laguerre polynomial
\begin{equation}
\frac{d}{dx}L^{(\alpha)}_{\mn}(x)=-L^{(\alpha+1)}_{\mn-1}(x)~,
\end{equation}
we obtain eq.(\ref{eq:a3vec-f}).

\section{Calculation of $\mathcal{A}_{3}$}
\label{sec:scal}

In this appendix we give the steps to obtain
eq.(\ref{eq:samp-3}).

Introducing a Grassmann parameter $\eta$,
we write the vertex operator
$\mathbb{V}_{A}(\xi,\theta;k_{M};\zeta_{M})$
in an exponential form \cite{IM}\cite{CIMM2}:
\begin{eqnarray}
\lefteqn{\mathbb{V}_{A}(\xi,\theta;k_{M};\zeta_{M})
=\int d\eta\; :\exp\left(
 i\sqrt{\frac{\alpha^{\prime}}{2}} k_{M}\mathbb{X}^{M}
 +\frac{i}{2}\eta\zeta_{M}\dot{\mathbb{X}}^{M}\right)
 (\xi,\theta):}\\
&& =\int d\eta:\exp\left(
 i\sqrt{\frac{\alpha^{\prime}}{2}} k_{\mu}\mathbb{X}^{\mu}
 +\frac{i}{2}\eta\zeta_{\mu}\dot{\mathbb{X}}^{\mu}\right):
 :\exp \left[ i\sqrt{\frac{\alpha^{\prime}}{2}}
     \left(\kappa\mathbb{Z}+\overline{\kappa}\overline{\mathbb{Z}}\right)
     +\frac{i}{2}\eta\left(e\dot{\mathbb{Z}}
             +\overline{e}\dot{\overline{\mathbb{Z}}}
     \right) \right]:\;.\nonumber
\end{eqnarray}
In a similar way to eq.(\ref{eq:2gauge}),
we factorize the correlation function in the integrand
on the r.h.s.\ in eq.(\ref{eq:samp-2})
into two pieces;
the piece depending on $\mathbb{X}^{\mu}$ $(\mu=0,\ldots,p)$
and that on $\mathbb{Z}$ and $\overline{\mathbb{Z}}$.

First we consider the former piece, namely the factor which depends on
$\mathbb{X}^{\mu}$.
We further divide this factor into the contribution from
the bosonic coordinates and that from the worldsheet fermions.
The bosonic contribution is the same as eq.(\ref{eq:tvt1})
with the replacement $a\rightarrow \frac{\eta\theta}
                      {\sqrt{2\alpha^{\prime}} }$.
The fermionic contribution becomes
\begin{eqnarray}
&&\hspace{-1em}\langle 0| :\frac{1}{2}\zeta^{(1)}_{\mu}
   (\psi^{\mu}+\widetilde{\psi}^{\mu})(-x^{(1)}):
 :\exp\left[-\theta^{(2)}\sqrt{\frac{\alpha^{\prime}}{2}}
              k_{\mu}^{(2)}(\psi^{\mu}+\widetilde{\psi}^{\mu})
   -\frac{1}{2}\eta\zeta^{(2)}_{\mu}(\psi^{\mu}+\widetilde{\psi}^{\mu})
   \right](-x^{(2)}):\nonumber\\
&& \hspace{10em}\times  :\frac{1}{2}\zeta^{(2)}_{\mu}
   (\psi^{\mu}+\widetilde{\psi}^{\mu})(-x^{(2)}):|0\rangle\nonumber\\
&&=-\frac{\zeta^{(1)}\pdot\zeta^{(2)}}{x^{(1)}-x^{(2)}}
 +\eta\theta^{(2)}\sqrt{2\alpha^{\prime}}
 \frac{ (k^{(2)}\pdot\zeta^{(1)})(\zeta^{(2)}\pdot\zeta^{(3)})
        -(k^{(2)}\pdot\zeta^{(3)})(\zeta^{(1)}\pdot\zeta^{(2)})}
      {(x^{(1)}-x^{(2)})(x^{(2)}-x^{(3)})}~.
\end{eqnarray}
Combining the results in the bosonic and the fermionic sectors
and using the polarization conditions for $\zeta^{(a)}$ $(a=1,2,3)$,
we find that the contribution from $\mathbb{X}^{\mu}$
to the correlation function in eq.(\ref{eq:samp-2}) is
\begin{eqnarray}
&&\prod_{\mu=0}^{p}\delta(k^{(1)}_{\mu}+k^{(2)}_{\mu}+k^{(3)}_{\mu})
\prod_{1\leq c<d\leq 3}\exp
\left[\frac{i}{2}\theta^{\mu\lambda}k^{(c)}_{\mu}k^{(d)}_{\lambda}
     \epsilon(\tau^{(c)}-\tau^{(d)})\right]
  (x^{(c)}-x^{(d)})^{2\alpha^{\prime}k^{(c)}\pdot k^{(d)}}
\nonumber\\
\hspace*{-1em}&&\hspace{-1em}
\times\Bigg[ -\frac{\zeta^{(3)}\pdot\zeta^{(1)}}{x^{(1)}-x^{(3)}}
-\eta\theta^{(2)}}{\sqrt{2\alpha^{\prime}}
\frac{(\zeta^{(3)}\pdot\zeta^{(1)})\,G^{Z\overline{Z}}
      (\kappa^{(2)}\overline{e}^{(2)}+\overline{\kappa}^{(2)}e^{(2)})}
     {(x^{(1)}-x^{(3)})(x^{(1)}-x^{(2)})}\nonumber\\
 &&+\eta\theta^{(2)}}{\sqrt{2\alpha^{\prime}}
   \frac{ (k^{(2)}\pdot\zeta^{(1)})(\zeta^{(2)}\pdot\zeta^{(3)})
          -(k^{(2)}\pdot\zeta^{(3)})(\zeta^{(1)}\pdot\zeta^{(2)})
          +(k^{(3)}\pdot\zeta^{(2)})(\zeta^{(3)}\pdot\zeta^{(1)})}
         {(x^{(1)}-x^{(2)})(x^{(2)}-x^{(3)})}\Bigg] .\nonumber\\
\end{eqnarray}

Next we consider the remaining piece, namely the factor which depends
on $\mathbb{Z}$ and $\overline{\mathbb{Z}}$,
in the correlation function in eq.(\ref{eq:samp-2}).
Sending $x^{(1)}\rightarrow\infty$ and $x^{(3)}\rightarrow 0$,
we find that this piece takes the form of
\begin{eqnarray}
&&\hspace{-1em}
\langle 0| U_{\mm}^{(-,-1)}(-x^{(1)})
:\exp \left[ i\sqrt{\frac{\alpha^{\prime}}{2}}
     \left(\kappa^{(2)}\mathbb{Z}+\overline{\kappa^{(2)}}
       \overline{\mathbb{Z}}\right)
     +\frac{i}{2}\eta\left(e^{(2)}\dot{\mathbb{Z}}
             +\overline{e}^{(2)}\dot{\overline{\mathbb{Z}}}
     \right) \right](x^{(2)},\theta^{(2)}):\nonumber\\
&&\hspace{3em}\times
     U_{\mn}^{(+,-1)}(-x^{(3)})|0
     \rangle \nonumber\\
&& =(x^{(1)})^{-(2\mm+1)(1-\nu)}
    (x^{(2)})^{-2\alpha^{\prime}G^{Z\overline{Z}}
               \kappa^{(2)}\overline{\kappa}^{(2)}}
\exp\left[ \mathcal{C}(\nu;k^{(2)}_{m})
    +\eta\theta^{(2)}\sqrt{2\alpha^{\prime}}
      \frac{G^{Z\overline{Z}} \overline{\kappa}^{(2)}e^{(2)}}
            {x^{(2)}}  \right]   \nonumber\\
&&\times\langle \sigma,s|(\alpha_{1-\nu})^{\mm}b_{\frac{1}{2}-\nu}  
\maru\exp \left[ i\sqrt{\frac{\alpha^{\prime}}{2}}
     \left(\kappa^{(2)}\mathbb{Z}+\overline{\kappa^{(2)}}
       \overline{\mathbb{Z}}\right)
     +\frac{i}{2}\eta\left(e^{(2)}\dot{\mathbb{Z}}
             +\overline{e}^{(2)}\dot{\overline{\mathbb{Z}}}
     \right) \right](-x^{(2)},\theta^{(2)})\maru
     \nonumber \\
&&\hspace{3em}\times 
  (\overline{\alpha}_{-1+\nu})^{\mn}\overline{b}_{-\frac{1}{2}+\nu}
     |\sigma,s\rangle   ,
 \label{eq:renosup}
\end{eqnarray}
with $x^{(1)}=\infty$ and $x^{(3)}=0$.
Here we have used the renormal ordering formula (\ref{eq:renormal}).
The contribution from the bosonic coordinates
$Z$ and $\overline{Z}$ to the correlation
function on the r.h.s.\ in eq.(\ref{eq:renosup})
has the form of eq.(\ref{eq:aVavec})
with the parameter $a$ replaced with
$\frac{\eta\theta^{(2)}}{\sqrt{2\alpha^{\prime}}}$.
By using the commutation relation (\ref{eq:com}),
we find that the contribution from the fermionic coordinates
becomes
\begin{eqnarray}
&&\hspace{-1em}
\langle s| b_{\frac{1}{2}-\nu}\maru\exp
  \left[-\left(\theta^{(2)}\sqrt{\frac{\alpha^{\prime}}{2}}\kappa^{(2)}
         +\eta\frac{e^{(2)}}{2}\right)
         (\Psi+\widetilde{\Psi})
        -\left(\theta^{(2)}\sqrt{\frac{\alpha^{\prime}}{2}}
          \overline{\kappa}^{(2)} +\eta\frac{\overline{e}^{(2)}}{2}
          \right)
      (\overline{\Psi}+\widetilde{\overline{\Psi}})
   \right]\maru\nonumber\\
&&\hspace{2em}\times \overline{b}_{-\frac{1}{2}+\nu}|s\rangle
 \ \ \ =\frac{2}{\varepsilon}\left\{1
   +\frac{\eta\theta^{(2)}}{\sqrt{2\alpha^{\prime}}}
   \frac{2\alpha^{\prime}G^{Z\overline{Z}}
    (\kappa^{(2)}\overline{e}^{(2)}-\overline{\kappa}^{(2)}e^{(2)})}
    {x^{(2)}}\right\}~.
\end{eqnarray}

Gathering all the results obtained above and using the physical state
conditions, we obtain the three point amplitude (\ref{eq:samp-3}).


\end{document}